\documentclass[onecolumn,floatfix]{emulateapj}

\usepackage{color}
\usepackage{graphics}
\usepackage{graphicx}
\usepackage{array}
\usepackage{verbatim}

\shorttitle{Disks around kicked black holes}
\shortauthors{Ponce, Faber, \& Lombardi}

\begin{document}

\title{Accretion disks around kicked black holes: Post-kick Dynamics}

\author{Marcelo Ponce}
\affil{Center for Computational Relativity and Gravitation, Rochester Institute of Technology, Rochester, NY 14623, USA}
\email{mponce@astro.rit.edu}
\author{Joshua A. Faber}
\affil{Center for Computational Relativity and Gravitation and School of Mathematical Sciences, Rochester Institute of Technology, Rochester, NY 14623, USA}
\email{jafsma@rit.edu}
\author{James C. Lombardi, Jr.}
\affil{Department of Physics, Allegheny College, Meadville, PA 16335, USA}
\email{jalombar@allegheny.edu}

\begin{abstract}

Numerical calculations of merging black hole binaries indicate that asymmetric emission of gravitational radiation can kick the merged black hole at up to thousands of km/s, and  a number of systems have been observed recently whose properties are consistent with an active galactic nucleus containing a supermassive black hole moving with substantial velocity with respect to its broader accretion disk.  We study here  the effect of an impulsive kick delivered to a black hole on the dynamical evolution of its accretion disk using a smoothed particle hydrodynamics code, focusing attention on the role played by the kick angle with respect to the orbital angular momentum vector of the pre-kicked disk.  We find that for more vertical kicks, for which the angle between the kick and the normal vector to the disk $\theta\lesssim 30^\circ$, a gap remains present in the inner disk, in accordance with the prediction from an analytic collisionless Keplerian disk model, while for more oblique kicks with $\theta\gtrsim 45^\circ$, matter rapidly accretes toward the black hole.  There is a systematic trend for higher potential luminosities for more oblique kick angles for a given black hole mass, disk mass and kick velocity, and we find large amplitude oscillations in time in the case of a kick oriented $60^\circ$ from the vertical.\\
\end{abstract}

\keywords{Black hole physics, Accretion, accretion disks, Hydrodynamics}

\section{Introduction}
In the past few years, a combination of surprising numerical studies and astronomical observations have indicated that asymmetric momentum losses in gravitational radiation from the mergers of  binary black holes (BH) may produce ``kicks'' of up to several thousand km/s.  For the mergers of supermassive black holes (SMBH) at the centers of galaxies, the kicks may be large enough to eject the remnant BH out of the galactic  center if not out of the galaxy entirely.  Kicked BHs may have already been observed indirectly as active galactic nuclei (AGN) with different components at different redshifts; broad-line regions that are thought to originate near the BH itself will remain bound to the kicked BH and acquire its new line-of-sight velocity, while narrowline systems that are produced further away will become unbound and remain behind.

While kicks from the mergers of unequal-mass, non-spinning BHs have long been predicted by post-Newtonian calculations \citep{Fitchett:1984qn}, the ability to evolve black holes in fully general relativistic simulations has considerably expanded our view.  Indeed, it is the merger of equal-mass, {\em spinning} BH that produce the largest kicks calculated to date, with maximum values of up to $4000$km/s possible for configurations with carefully chosen alignments \citep{Campanelli:2007ew,Campanelli:2007cga,Gonzalez:2007hi}.  While speeds this large represent only a small fraction of the likely merging BH parameter space, Monte Carlo simulations of merging BHs with arbitrary spins of dimensionless magnitude $S/M^2=0.97$ find that the mean kick for  BH systems with mass ratios uniformly distributed between $0$ and $1$ is 630km/s, with more than 20\% of the kicks larger than 1000km/s \citep{Lousto:2009ka}.   Such kicks have the potential to unbind the remnant from smaller galaxies, or displace the BH and any bound gas for larger galaxies.  In roughly half of all major mergers between comparable-mass SMBHs, the merged SMBH should remain displaced for 30 Myr outside the central torus of material that would power an AGN \citep{Komossa:2008as}.

Somewhat more recently, several AGN have been observed to contain broadline emission systems that appear to have very different redshifts than the narrowline emission systems.   In the first of these, SDSSJ092712.65+294344.0, a blueshift of $2650$km/s is observed for the broadline systems relative to narrow lines \citep{Komossa:2008qd}.  Although several different physical models have been proposed to account for the observations, including broadline emission from the smaller SMBH within a binary surrounded by a disk \citep{Dotti:2008yb,Bogdanovic:2008uz}, a pair of SMBHs at the respective centers of interacting galaxies \citep{Heckman:2008en}, and even spatial coincidence \citep{Shields:2008kn}, the kicked disk model remains entirely plausible.  Since then, several more systems with similar velocity offsets between different AGN emission regions have been discovered, including SDSS J105041.35+345631.3 [3500km/s; \citep{Shields:2009jf}], CXOC J100043.1+020637 [1200km/s; \citep{Civano:2010es}], and E1821+643 [2100 km/s; \citep{Robinson:2010ui}].

Given a number of recent results that suggest that binary SMBH spins should align with the angular momentum of the accretion disk prior to merger,
these large kick velocities are somewhat unexpected.
Indeed, aligned spins can produce kicks of several hundred km/s \citep{Campanelli:2007cga} but not the ``superkicks'' with recoil velocities $\gtrsim 1000$km/s.
\citet{Bogdanovic:2007hp} suggest  that the accretion torques in gas-rich (``wet'') mergers should suffice to align the SMBH spins with the accretion disk prior to merger.
In \citet{Dotti:2009vv}, high resolution simulations of SMBH binaries whose orbits counter-rotate with regard to a surrounding disk indicate that they should undergo an angular momentum flip long before merger.
With typical spin-orbit misalignments of no more than $10^\circ-30^\circ$ depending on the parameters of the disk, in particular its temperature, they find that the recoil kick during mergers should have a median value of $70$km/s, with superkicks an exceedingly rare event \citep{Dotti:2009vz}.

While some of the observed candidate recoil velocities are so large that they fall well out into the tail of the kick velocity probability distribution function, it is difficult to constrain exactly how many systems indicate potential recoils given the challenges in clearly distinguishing multiple velocity components within a single AGN.  Given this difficulty, we attempt here to study potential electromagnetic (EM) signatures that would originate from post-merger disks.

The qualitative details of pre-merger evolution have been studied by other groups, and a relatively coherent picture emerges.  In general, each SMBH may be surrounded by a circum-BH accretion disk extending out to a distance substantially smaller than the binary separation, since orbits near the outer edge are unstable to perturbations from the other SMBH.  A circumbinary disk may be present as well, extending inward to a distance of roughly a few times that of the binary separation, with a gap appearing between the circumbinary disks and the inner disks.  Previous hydrodynamical studies have shown that  the inner edge of the circumbinary disk is driven into a high-eccentricity configuration that precesses slowly, while a 2-armed spiral density wave is formed extending out to larger radii \citep{Macfadyen:2006jx}.
Meanwhile, the inner disks will be fed by mass transfer from the circumbinary disk \citep{Hayasaki:2006fq,Kimitake:2007fs}, as the $m=2$ azimuthal gravitational perturbation induces an elongation in the outer disk.  For circular orbits, the mass transfer rate is relatively constant, while for elliptic binaries the mass transfer takes on a more periodic character.  Finally, once the binary begins its gravitational radiation-driven plunge, the binary decouples form the outer disk, and mass transfer basically ceases.

Among the first predictions of the observational consequences of a post-kick accretion disk are those in \citet{Loeb:2007wz}, who find that off-center quasars could be observed for up to $10^7$ years after a SMBH merger. 
\citet{Schnittman:2008ez} find that the inner edge of the circumbinary disk is likely to occur at roughly 1000$M$, in typical relativistic units where $G=c=1$, with the value only weakly dependent on the typical disk parameters like the assumed $\alpha$-parameter.  Based on semi-analytic models of the post-merger accretion disk, they find the potential for observable infrared emission lasting hundreds of thousands of years, leading to the prediction that several such sources might be present today, though they would be difficult to disentangle from other AGN sources.    On much shorter timescales immediately prior to the merger, the dissipation of gravitational radiation energy through spacetime metric-induced shearing of the disk can also lead to enhanced emission in the optically thin components of the disk \citep{Kocsis:2008va}.  

One of the first studies of post-kick disk dynamics was performed in \citet{Shields:2008va}, who analyzed the approximate physical scales characterizing the post-merger disk and concluded that the total energy available to be dissipated by shocks is roughly $\frac{1}{2}M_b v_{\rm kick}^2$, where $M_b$ is the mass of the portion of the disk that remained bound and $v_{\rm kick}$ is the SMBH kick velocity.  They estimated from a simple analytical model that an excess luminosity of $\sim 10^{46}$ erg/s would be observable for roughly 3000 years, with a characteristic observed temperature of $\sim 10^7$K assuming $v_{\rm kick}=1000$km/s.  This model is checked by means of a {\em collisionless} N-body simulation of a disk around a kicked SMBH, which found rough agreement with the analytical model and confirmed the prediction of a visible soft X-ray flare that would last for a few thousand years after the initial kick.

Using a 2-d version of the FLASH code, \citet{Corrales:2009nv} are  able to study the response of a thin disk if the kick is  directed in the equatorial plane.  They find  that the characteristic response is  a one-armed spiral shock wave, capable of producing total luminosities up to $\sim 10\%$ of the Eddington luminosity on a timescale of months to years.  The relativistic decrease of the total SMBH mass, attributable to the energy carried off in gravitational waves and roughly $5\%$ of the original total, leads  to a decrease in the luminosity of approximately $15\%$ but does  not provide  a clear signature that can  be disentangled from the other global processes occurring in the disk.  Nevertheless, with future X-ray instruments such as the Square Kilometer Array,
impulsive changes to a disk may be observable in the jet emission  \citep{O'Neill:2008dg}. 
 Even in the case where the BH kick is  very small, the secular, as opposed to dynamical, filling of the inner region of the disk should produce an afterglow that could exceed the Eddington luminosity if the accretion rate is  sufficiently large prior to the binary decoupling and merger \citep{Shapiro:2009uy}. 

The most direct comparison to the calculations we present here is found in \citet{Rossi:2009nk}.  Using an analytic treatment of disks with power-law density profiles, they construct a model for disk evolutions in which, immediately after the kick, fluid elements are assumed to circularize at radii determined by their specific angular momentum, with the resulting energy gain by shocks released as EM radiation.  Their work establishes that the primary energy reservoir for kicked disks  is potential energy that can be released by elliptical orbits, not the relative kinetic energy of the kick itself nor the impulse sent through the disk by instantaneous mass loss to gravitational radiation by the central SMBH during the merger.  They also perform detailed 2-d Eulerian and 3-d SPH simulations of post-kick accretion disks, although there are some important differences between the latter simulations and most of our SPH studies, as we discuss throughout the paper below.   Among these, they assume that disks are isothermal, with all shock heating immediately radiated away, whereas we evolve the energy equation and allow the fluid to heat.  They also include an ``accretion radius'', $R_{acc}$, such that particles that approach closer than that distance to the black hole are accreted and removed from the simulation domain (G. Lodato, private communication), whereas all particles remain in our simulation throughout the duration of a run.  As such, their results and ours  bracket the range of possible heating scenarios.  

The 2-d calculations by \citet{Rossi:2009nk} of razor-thin disks using the ZEUS code indicate that vertical kicks (that is, in the direction of the disk angular momentum vector) lead to modulated emission, unlike all other kick angles they consider. This result is confirmed by their 3-d SPH simulations. In-plane kicks develop a clear spiral-wave structure with accretion streams forming as the simulation progressed, but a smooth luminosity profile.  Their 3-d SPH simulations with oblique (less vertical) kicks  indicate essentially a 2-phase model for the observed EM emission.  Immediately after the kick, the majority of the luminosity may be attributed to the innermost region of the disk dissipating the kinetic energy it acquired during the kick, while at later times, after roughly one standard timescale of the disk ($\hat{t}=1$ in the notation of Eq.~\ref{eq:tscale} below) and thereafter, the dominant dissipation mode is potential energy from infalling material on elliptical orbits.  Post-kick disks are found to be rather compact, extending out to roughly the ``bound radius'' ($\hat{r}=1$ according to the notation of our Eq.~\ref{eq:rscale}) with steep density dropoffs at larger radii.  Luminosities are generally highest for in-plane kicks, with roughly a factor of four difference in peak luminosity between largest peak luminosity (in-plane kick) and the smallest (vertical kick), with peaks occurring later for more oblique kicks.  Relativistic BH mass decrease was found to be unimportant at large radii, and potentially important only in the vicinity of the BH (out to a few hundred Schwarzschild radii) where the effect of the kick is merely perturbative compared to the nearly relativistic Keplerian velocities.

Numerical relativity groups have also considered the hydrodynamics of matter around both binary SMBHs and kicked BHs, typically at much smaller size scales.  In \citet{Bode:2009mt}, flows of gas  around a binary SMBH system are  considered at scales roughly $10^5$ smaller than typical Newtonian calculations, spanning scales roughly 1AU across, rather than $\sim$parsec scales.  They find  that EM emission is  dominated by variability created by Doppler beaming of the SMBHs as they shock the gas, leading to an EM signal that demonstrates the same periodicity as the gravitational wave signal, with corresponding peaks in the timing of the maximum emission for each.  In a follow-up work, Bode and collaborators \citep{Bode:2011tq} predict that observable EM emission from near the SMBHs is much more likely to arise in a hot accretion flow, in which a flare would be seen coincident with the merger.
 In \citet{Megevand:2009yx},  the effect of a kicked BH moving through the equatorial plane of an accretion torus is  considered using a fully general relativistic Eulerian hydrodynamics code.  In their simulations, the newly-merged SMBH is  surrounded by a torus extending out to 50M (50AU for a $10^8M_\odot$ SMBH),  and the overall timescale studied is  approximately 10,000M (in relativistic units with $G=c=1$), or about two months.  They find  that a kick in the direction of the equatorial plane of the torus produces
the strongest shock in the system and therefore the strongest EM emission, consistent with studies that examine disks on substantially larger scales.  By ray-tracing their simulations in a post-processing step \citep{Anderson:2009fa}, they confirm that simple Bremsstrahlung luminosity estimates yield a qualitatively accurate picture of the disk luminosity for high-energy radiation, while their torus is optically thick to low-energy emission.
 
Numerical calculations of vacuum EM fields surrounding an SMBH merger indicate that they could contribute to periodic emission \citep{Palenzuela:2009yr,Palenzuela:2009hx} but are likely to be too small in amplitude and at the wrong frequencies ($\sim 10^{-4}$Hz) to be observed directly \citep{Mosta:2009rr}.  Such mergers could produce observable levels of Poynting flux in jets \citep{Palenzuela:2010nf}, however, through a binary analogue of the Blandford-Znajek mechanism \citep{Blandford:1977ds}, which seems especially effective for spinning BHs \citep{Neilsen:2010ax}.
Calculations of non-Keplerian accretion disks in general relativity indicate  that the spiral wave structures seen in Newtonian simulations could exist in relativistic models with small disks when  the BH kick is  sufficiently small, but that larger kicks disrupt the spiral pattern, as could dissipative processes such as magnetic stress or radiative cooling \citep{Zanotti:2010xs}.
 The inferred emission due to synchrotron emission from a relativistic disk is  considered by the Illinois group \citep{Farris:2009mt,Farris:2011vx}, who find  that emission could peak at a luminosity of $\sim 10^{46}$ erg/s, a few orders of magnitude brighter than the corresponding bremsstrahlung luminosity and potentially observable by either WFIRST or LSST.

The outline of the paper is as follows.  In Section~\ref{sec:scales}, 
we introduce the physical scales that define the kicked disk problem, and discuss  the basic dynamics of disks around a kicked black hole when hydrodynamic effects are ignored.  In Sec.~\ref{sec:SPH}, we describe the SPH code used to perform 3-dimensional  simulations, including versions that seek to replicate previous results and those that use new techniques to bracket the range of physical physical predictions.   Results from these simulations, focusing on the hydrodynamic and thermodynamic evolution of the disk, are reported in Sec.~\ref{sec:runs}.  Finally, in Sec.~\ref{sec:discussion}, we lay out consequences of our results and plans to extend them in the future.

\section{Physical scales and qualitative expectations}\label{sec:scales}

\subsection{Physical scales}\label{subsec:scales}
Throughout this paper, we use scaled units, denoted by hats, under the assumption that $G=M_{\rm BH}=v_{\rm kick}=1$.  A single unit of time, for example, is thus $GM_{\rm BH}v_{\rm kick}^{-3}$.  Choosing reference values $M_{\rm BH}=10^8 M_\odot$ and $v_{\rm kick}=10^8~{\rm cm/s}$, the resulting time and distance scales are defined as
\begin{eqnarray}
t&=& 1.327\times 10^{10} \left(\frac{M_{\rm BH}}{10^8M_\odot}\right) \left(\frac{v_{\rm kick}}{10^8~{\rm cm/s}}\right)^{-3}\hat{t}~{\rm s} =421 \left(\frac{M_{\rm BH}}{10^8M_\odot}\right) \left(\frac{v_{\rm kick}}{10^8~{\rm cm/s}}\right)^{-3}\hat{t}~{\rm yr} \label{eq:tscale}\\
d&=& 1.327\times 10^{18} \left(\frac{M_{\rm BH}}{10^8M_\odot}\right) \left(\frac{v_{\rm kick}}{10^8~{\rm cm/s}}\right)^{-2}\hat{d}~{\rm cm} = 8.87\times 10^4\left(\frac{M_{\rm BH}}{10^8M_\odot}\right) \left(\frac{v_{\rm kick}}{10^8~{\rm cm/s}}\right)^{-2}\hat{d}~{\rm AU}\nonumber\\
&=&0.43\left(\frac{M_{\rm BH}}{10^8M_\odot}\right) \left(\frac{v_{\rm kick}}{10^8~{\rm cm/s}}\right)^{-2}\hat{d}~{\rm pc}.\label{eq:rscale}
\end{eqnarray}
Because  the disk is evolved without self-gravity, the disk mass scale is formally independent of the BH mass that sets the rest of the physical scales.

For a disk of total mass $m_{\rm disk}$, we may define a set of quantities marked with tildes by choosing a reference value $m_{\rm disk}=10^4M_\odot$.  The physical scales for energy, its time derivative, volume density, and surface density are then given respectively by 
\begin{eqnarray*}
E &=& 1.99\times 10^{53}\left(\frac{m_{\rm disk}}{10^4~M_\odot}\right) \left(\frac{v_{\rm kick}}{10^8~{\rm cm/s}}\right)^2 \tilde{E}~{\rm erg}\\
\frac{dE}{dt} &=& 1.50\times 10^{43}\left(\frac{m_{\rm disk}}{10^4~M_\odot}\right)\left(\frac{M_{\rm BH}}{10^8M_\odot}\right)^{-1} \left(\frac{v_{\rm kick}}{10^8~{\rm cm/s}}\right)^{5}\frac{d\tilde{E}}{d\hat{t}}~{\rm erg/s}\\
\rho&=&8.51\times 10^{-18}~\left(\frac{m_{\rm disk}}{10^4~M_\odot}\right)\left(\frac{M_{\rm BH}}{10^8M_\odot}\right)^{-3}\left(\frac{v_{\rm kick}}{10^8~{\rm cm/s}}\right)^6\tilde{\rho}~{\rm g/cm^3}\\
&=&1.26\times 10^5~\left(\frac{m_{\rm disk}}{10^4~M_\odot}\right)\left(\frac{M_{\rm BH}}{10^8M_\odot}\right)^{-3}\left(\frac{v_{\rm kick}}{10^8~{\rm cm/s}}\right)^6\tilde{\rho}~{\rm M_\odot/pc^3}\\
\Sigma &=& 11.3~\left(\frac{m_{\rm disk}}{10^4~M_\odot}\right)\left(\frac{M_{\rm BH}}{10^8M_\odot}\right)^{-2}\left(\frac{v_{\rm kick}}{10^8~{\rm cm/s}}\right)^4\tilde{\Sigma}~{\rm g/cm^2}. 
\end{eqnarray*}
Assuming the fluid is a fully ionized ideal gas with mass fractions $X=0.7,~Y=0.28$, the mean molecular mass is $\mu=2/(1+3X+0.5Y)=0.617$, and the characteristic temperature scales like
\begin{eqnarray*}
T = \frac{\mu m_p v_{\rm kick}^2}{k_B}\tilde{T} = 7.48\times 10^7 \left(\frac{v_{\rm kick}}{10^8~{\rm cm/s}}\right)^2\tilde{T}~{\rm K}, 
\end{eqnarray*}
where $m_p$ is the mass of a proton, and $k_B$ is the Boltzmann constant.  
The optical depth for Thomson scattering, $\tau\equiv \kappa_e\Sigma$, where $\kappa_e \approx 0.2(1+X)=0.34~{\rm cm^2/g}$ is appropriate for non-relativistic
plasma, is given by
\begin{equation}
\tau\equiv \kappa_e\Sigma \approx 3.8~\left(\frac{m_{\rm disk}}{10^4~M_\odot}\right)\left(\frac{M_{\rm BH}}{10^8M_\odot}\right)^{-2}\left(\frac{v_{\rm kick}}{10^8~{\rm cm/s}}\right)^4 \tilde{\Sigma}. \label{eq:depth}
\end{equation}

\subsection{Qualitative expectations}\label{sec:collisionless}

The simplest model for a disk around a merging SMBH binary consists of a 2-dimensional, infinitely thin disk with a perfectly Keplerian rotational profile.  If we assume that Newtonian gravity applies, we have a nearly scale-free system, where only the mass of the SMBH binary contains units.  In what follows, we work in dimensionless units such that $G=M_{\rm BH}=1$, where $M_{\rm BH}$ is the total mass of the SMBH binary, assumed for the moment to be equal to the total mass of the merged SMBH that will be formed in the merger and immediately kicked with velocity $v_{\rm kick}$ at an angle $\theta$ relative to the angular momentum of the 2-d disk.  We also choose $v_{\rm kick}=1$ to set the overall scaling for the remaining quantities we consider, denoting all quantities in these units by hats.  The Keplerian rotational velocity, for instance, is given by $\hat{v}_0=\hat{r}^{-1/2}$.   Note that our unit system results in the speed of light {\em not} being set to unity.  Note that our conventions are very similar to those found in \citet{Rossi:2009nk}, except for the definition of the kick angle: we define $\theta$ to be the angle away from vertical, whereas they define $\theta$ to be an the angle between the kick and the initial disk plane.

In what follows, we assume that the disk orbits in the $x-y$ plane, with angular momentum in the positive $z$-direction.  We define $\phi$ to be the azimuthal angle in this plane, measured counterclockwise from the $x$ axis.  The BH kick falls in the $x-z$ plane, and we assume that the SMBH proceeds to move off with constant velocity, feeling no accelerations from either the disk or the galactic potential on the timescales of interest.

The relative velocity and specific energy of points in the disk with respect to the newly kicked BH is given by
\begin{eqnarray*}
\hat{v}(r,\phi)&\equiv&\hat{v}_0-\hat{v}_{\rm k}=(-\sin\theta-\hat{r}^{-1/2}\sin\phi,~\hat{r}^{-1/2}\cos\phi,~-\cos\theta)\\
\hat{E}(r,\phi)&=&\frac{1+2\hat{v}_0\sin\theta\sin\phi-\hat{v}_0^2}{2}=\frac{1+2\hat{r}^{-1/2}\sin\theta\sin\phi-\hat{r}^{-1}}{2}. 
\end{eqnarray*}
Accordingly, the critical radius $\hat{r}_b$ as a function of azimuthal angle $\phi$  inside of which the matter is bound to the kicked BH and unbound outside (cf. Fig.~1 of \cite{Rossi:2009nk}) is given by
\begin{equation}
\hat{r}_b=\hat{v}_{\rm b}^{-2}=\left(\sin\theta\sin\phi+\sqrt{1+\sin^2\theta\sin^2\phi}\right)^{-2}.
\end{equation}

\begin{figure}
        \includegraphics[width=\columnwidth]{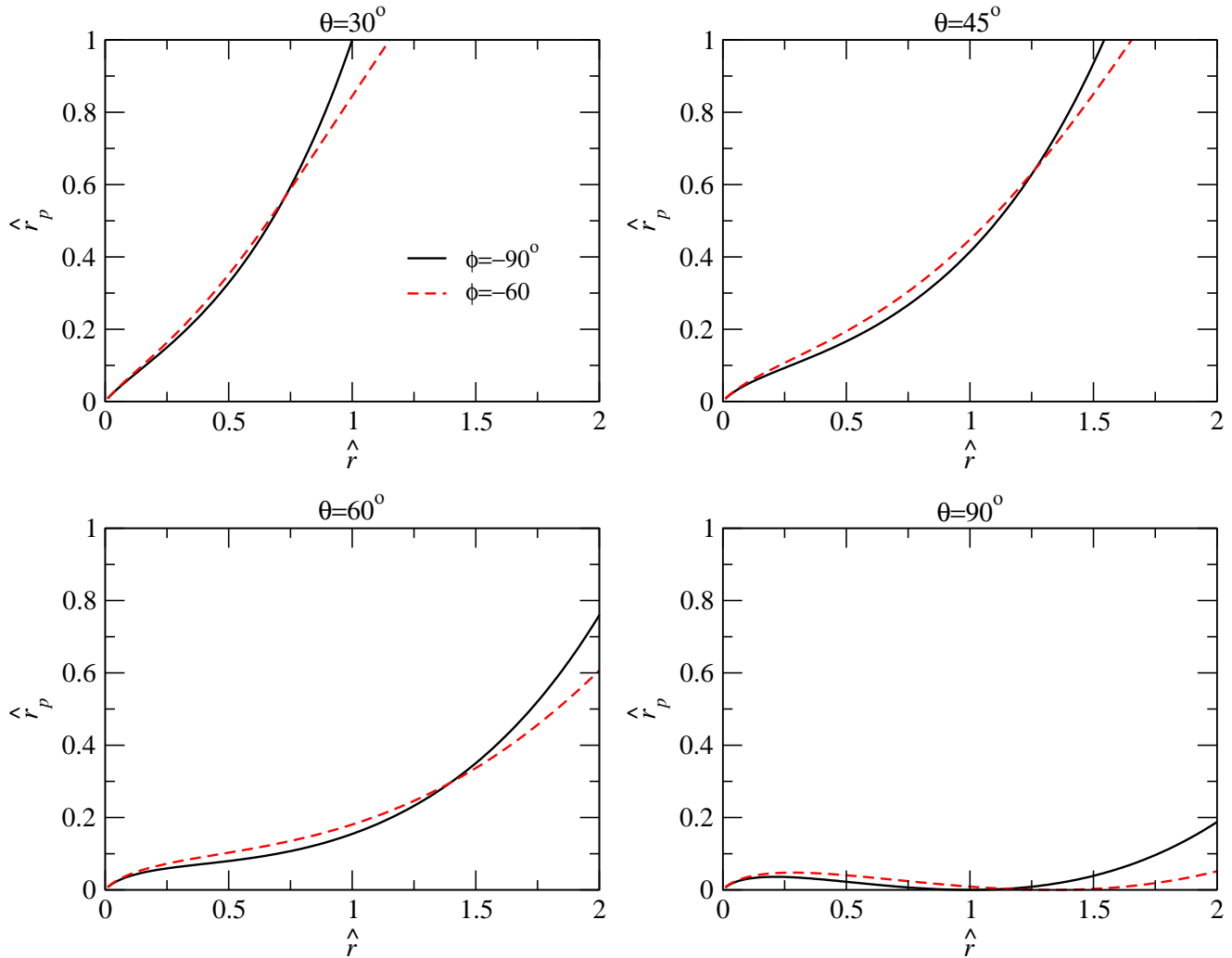}
	\caption{Particle periastron values after the BH kick as a function of radius for azimuthal angles $\phi=-60^\circ,~-90^\circ$ and selected  kick angles $\theta$.  Assuming the disk has an inner edge at $\hat{r}=\hat{r}_0$ prior to the kick, the minimum value of $\hat{r}_p$ decreases as $\theta$ increases from $0^\circ$, a vertical kick, to $\theta=90^\circ$, an in-plane kick, with particularly rapid changes for large values of $\theta$.}
	\label{fig:rperi}
\end{figure}

For both bound and unbound particles, we may calculate the full trajectory of the particles assuming they remain collisionless by solving the two-body problem with respect to the kicked black hole.  
Of particular importance for our later discussion in the post-kick periastron radius,
$\hat{r}_p$, shown in Fig.~\ref{fig:rperi},
for longitudes  $\phi=-60^\circ$ and $-90^\circ$ where the values can be quite small, indicating significant potential mass flow toward the BH. 
 Values are systematically smaller for kick angles closest to the original disk plane ($\theta=90^\circ$), and,  if we impose an inner edge on the original disk,  there is a minimum periastron for all post-kick particles that decreases dramatically as the kick angle becomes more in-plane.  This leads to a testable prediction  for 3-d dynamical simulations: a ``gap'' should be present for more vertical kick angles, disappearing only once collisions in the inner region of the post-kick disk facilitate the angular momentum loss required to feed flow toward the SMBH, as noted for in-plane and vertical kicks in \cite{Rossi:2009nk}.   For vertical kicks, 
we expect that it should take several times the orbital period at the inner edge of a disk before any significant amount of matter is present near the SMBH. For in-plane kicks, the filling of the inner disk should be much more rapid.


\section{Smoothed Particle Hydrodynamics for kicked accretion disks}\label{sec:SPH}

\subsection{Initial data}

Following the methods outlined in \citet{RubioHerrera:2004pf,RubioHerrera:2005fv} and similar works, we construct semi-analytic models of accretion disks in hydrodynamic equilibrium to use as initial conditions  before laying down particles using a Monte Carlo technique.
 To do so, we first assume a that the orbital velocity is independent of the height within the disk and varies only with cylindrical radius. 
Integrating the force equation for a system in stationary equilibrium,
\begin{equation}
-\frac{\nabla P}{\rho}+\nabla\left(\frac{GM}{r}\right)=-\frac{l(r_c)^2}{r_c^3}\hat{\bf{r_c}}, 
\end{equation}
and assuming the pressure $P=k\rho^\gamma$, where $\gamma=5/3$ is the adiabatic index of the fluid, $M$ the mass of the central BH, $r$ and $r_c$ are the spherical and cylindrical radii, $\hat{\bf{r_c}}$ is the unit vector associated with the cylindrical radius, and $l(r_c)$ is the chosen 
radial specific angular momentum profile, we find
\begin{equation}
\eta\equiv \frac{\gamma}{\gamma-1}\frac{P}{\rho}=\frac{GM}{r}+\int \frac{l(r_c)^2}{r_c^3}dr_c-K, \label{integratedForceEquation}
\end{equation}
where $K$ is an integration constant.  For the Keplerian profile $l(r_c)=\sqrt{GMr_c}$, the integral on the right hand side of eq.\ (\ref{integratedForceEquation}) evaluates to $-GM/r_c$, and the disk can exist only in the $z=0$ plane (where $r=r_c$).

To get a disk with finite extent in the radial and vertical directions, we choose a non-Keplerian rotation profile.  Here, we assume a power-law form, noting that it must satisfy $l(r)\propto r^\kappa$ with $\kappa<0.5$ to yield a compactly bounded configuration.  We define rotation parameters through the relation
\begin{equation}
\int \frac{l(r_c)^2}{r_c^3} dr_c = -cr_c^\alpha\label{eq:rotation}
\end{equation}
and find that the top edge of the disk, where $\eta=0$, yields the condition that 
\begin{equation}
z(r_c)=\sqrt{\left(\frac{GM}{cr_c^\alpha+K}\right)^2-r_c^2}.
\end{equation}

Assuming (hatted) units such that $G=M=1$, the inner and outer edges of the disk for a sub-Keplerian rotation profile ($\kappa<0.5$ and thus $\alpha<-1$) are given by the two real roots of the equation $\hat{z}(\hat{r})=0$, or
\begin{equation}
\hat{c}\hat{r}^{\alpha+1}+\hat{K}\hat{r}=1.
\end{equation}

To fix the inner and outer radii at $\hat{r}_i$ and $\hat{r}_0$ respectively, we determine $\hat{c}(\alpha)$ and $\hat{K}(\alpha)$ as follows.  Defining
$\hat{R}_i=\hat{r}_i^{\alpha+1}$ and $\hat{R}_o=\hat{r}_o^{\alpha+1}$, we find
\begin{eqnarray*}
\hat{c}&=&\frac{\hat{r}_o-\hat{r}_i}{\hat{R}_i\hat{r}_o-\hat{R}_0\hat{r}_i}\\
\hat{K}&=&\frac{\hat{R}_i-\hat{R}_o}{\hat{R}_i\hat{r}_o-\hat{R}_o\hat{r}_i}.
\end{eqnarray*}
For the case $\alpha=-2$, corresponding to a constant $l(r_c)$, the solution is easy to state explicitly: noting that $\hat{R}_o=1/\hat{r}_o$ and $\hat{R}_i=1/\hat{r}_i$, we find
\begin{eqnarray*}
\hat{c}&=&\frac{\hat{r}_o-\hat{r}_i}{\hat{r}_o/\hat{r}_i-\hat{r}_i/\hat{r}_o}=\frac{\hat{r}_i\hat{r}_o}{\hat{r}_o+\hat{r}_i}\\
\hat{K}&=&\frac{1/\hat{r}_i-1/\hat{r}_o}{\hat{r}_o/\hat{r}_i-\hat{r}_i/\hat{r}_o}=\frac{1}{\hat{r}_o+\hat{r}_i}.
\end{eqnarray*}

In general, the easiest method to achieve a specific disk height $\hat{z}_{max}$ is to vary $\alpha$ and check numerically where the disk reaches its maximum height, iterating until the correct value is achieved.
Relatively thin disks for which the radial extent is significantly greater than the vertical tend to be nearly 
Keplerian, with $\alpha=-1-\epsilon_\alpha$, where the $\epsilon_\alpha$ is positive and $\ll 1$.  

\begin{figure}
\begin{center}
\includegraphics[width=0.8\columnwidth]{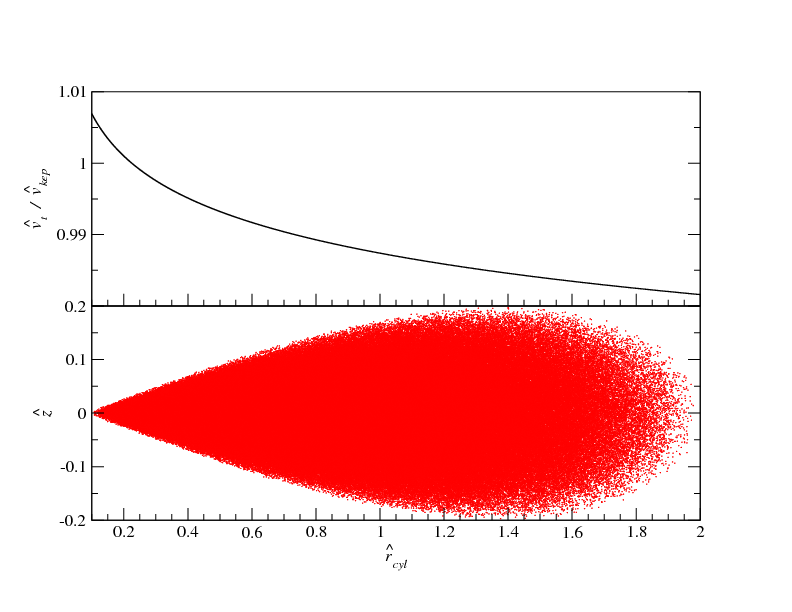}
\caption{Top: The rotational profile of the disk, expressed as a ratio of the actual tangential velocity to the the Keplerian velocity, shown as a function of radius.  Bottom: Initial disk configuration, projected onto the $\hat{r}_c -\hat{z}$ plane.  It has an inner radius $\hat{r}_i=0.1$, outer radius $\hat{r}_o=2.0$, and maximum height $\hat{z}_{max}=0.2$.    Finite disks require super-Keplerian rotation on the inner regions and sub-Keplerian rotation outside so that the centripetal acceleration respects the additional pressure force component.  The density maximum occurs at $\hat{r}=0.23$, where $\hat{v}_t=\hat{v}_{kep}$.}
\label{fig:diskt0}
\end{center}
\end{figure}

For all the runs shown below, we chose initial parameters $\hat{r}_i=0.1$, $\hat{r}_o=2.0$, and $\hat{z}_{max}=0.2$ for our initial disk, resulting in an SPH discretization and rotation curve we show in Fig.~\ref{fig:diskt0}.  This corresponds to a choice of parameters $\hat{c}=0.9584$, $\alpha=-1.017$, and $\hat{K}=2.651\times10^{-2}$.
Once we specify the adiabatic index $\gamma=5/3$, we are left with a free parameter in the adiabatic constant $k\equiv P/\rho^\gamma$.  Varying $k$ has the effect of rescaling the density (see eq.\ \ref{integratedForceEquation}), and thus allows us to adjust the overall disk mass while leaving a uniform specific entropy. 

\subsection{Hydrodynamical Evolution}\label{sec:evoleqs}

The code used to perform our calculations combines the parallelization found in the {\tt StarCrash} SPH code with a number of refinements described in \citet{Lombardi:2005nm} and \citet{Gaburov:2009kg}, which we summarize in an appendix.  Among these, we implement variable smoothing lengths for all of our particles following the formalism described in \citet{Springel:2001qb,Monaghan:2001MNRAS.328..381M,Price:2006iz}, which can be derived consistently from a particle-based Lagrangian.  Gravitational softening for particles near the BH is implemented in a self-consistent manner, similar to the method used in \cite{Price:2006iz}, which is critical to avoid individual and collective particle numerical instabilities for matter in the vicinity of the BH, as we discuss below.

Artificial viscosity is included in all runs using a standard ``Balsara switch'' \citep{Balsara:1995JCoPh.121..357B}.  In most of our runs, we implement the standard thermodynamic energy equation and allow the fluid to heat.  Such cases provide a minimum estimate for  the luminosity that can be produced by a kicked disk, since thermal pressure will damp continued shocking within the disk by lowering the ambient densities.  Since this ignores the effcets of radiative cooling losses, we also consider the more common approximation made in numerical simulations (see, e.g. \cite{Corrales:2009nv,Rossi:2009nk}, that all shock-generated energy is immediately radiated away, resulting in an adiabatic evolution (or isothermal, depnding on the initial hydrodynamic model).  Details of the implementation are discussed in the Appendix (see Eq.~\ref{eq:eintimplied}).  While it is relatively straightforward to calculate the luminosity in such a model, the accretion of large numbers of particles very near the BH is an inevitable consequence, and further code modifications are required to allow numerical evolutions, particularly the introduction of an accretion radius $R_{acc}$, such that particles that fall to smaller radii are assumed to be accreted by the BH and removed from the simulation.  We discuss this in much more detail in Sec.~\ref{sec:noheat}.



For a monatomic ideal gas, we can calculate the SPH internal energy as
\begin{equation}
E_{\rm INT}  = \sum_i \frac{1}{\gamma-1} m_i A_i \rho_i^{\gamma-1} = \frac{3}{2}\sum_i m_i A_i \rho_i^{2/3}. \label{eq:eint}
\end{equation}
Furthermore, the temperature is related to the specific internal energy $u_i = \frac{3}{2}A_i \rho_i^{2/3}$ via the ideal gas law $E_{\rm INT} =\frac{3}{2}k_BT$:
\begin{equation}
T_i=\frac{2}{3k_B} \mu m_p u_i = \frac{\mu m_p}{k_B}A_i \rho_i^{2/3} =\frac{\mu m_p}{k_B}\frac{P_i}{\rho_i},
\end{equation}
where $m_p$ is the mass of a proton and
$\mu=0.617$ is the mean molecular weight, assuming that the disk is a plasma with mass fractions $X=0.7,~Y=0.28$.

\section{SPH simulations}\label{sec:runs}

In this section, we present dynamical calculations of kicked disks.
The calculations of Sec.~\ref{subsec:heat} treat the dynamics assuming the gas is optically thick, so that radiative cooling can be neglected, while the calculations of Sec.~\ref{sec:noheat} assume that any heating that would have resulted from shocks is immediately radiated away.  The dynamics of real disks will, depending on the system parameters, fall somewhere between these two extremes and likely harbor a rich dependence on position and time.

\subsection{Calculations with Shock Heating}\label{subsec:heat}

To study the effect of the SMBH kick angle on the resulting disk evolution, 
we perform runs where the kick angles away from vertical are $15^\circ,~30^\circ,~45^\circ$, and $60^\circ$.
We have also performed a number of test calculations to optimize various SPH-related parameters including the number of neighbors and the Courant factors (see Sec.~\ref{sec:SPH}), along with numerical convergence tests to guarantee the validity of our simulations and determine the parameters for our fiducial production runs. 

Our production runs are summarized in Table~\ref{table:runs} below.  
The bound mass at the end of the simulation, $\tilde{M}_b$ , is defined by Eq.~\protect\ref{eq:bound}.  The angular momentum vector $\vec{L}_b$ of the bound matter is in the original coordinates, with the initial angular momentum of the disk in the $z$-direction.  The $\vec{L}_b$ vector defines the $z'$-direction used to construct cylindrical coordinates in the plots below.  The tilt angle of the kicked disk is defined by the condition $\theta_{tilt}\equiv\arccos (\vec{L}_{z;b}/|L_b|)$.  The approximate maximum luminosity of the disk may be estimated from $\left(\frac{dE_{int}}{dt}\right)_{max}$, though we note that we allow for the SPH particles to shock heat. 

Each run uses $N=5\times 10^5$ equal-mass SPH particles, and the number of neighbors is chosen to be $200$ initially.  All runs are started from the same relaxed disk configuration.  To construct it, we lay down particles uniformly in space and use the local density as the basis for a Monte Carlo rejection technique. This configuration is relaxed for a time interval $\hat{t}=160$, during which we apply a drag force 
\begin{equation}
a_{drag} = (v-v_R)/t_{rel}
\end{equation}
with $\hat{t}_{rel}=0.8$ as the chosen relaxation timescale and $v_R$ the exact rotation law satisfying Eq.~\ref{eq:rotation} above, with $v_R=l(r_c)/r_c$.

\begin{table}[ht!]
\caption{Summary of the production runs performed.}\label{table:runs}
\begin{center}
\begin{tabular}{cccccc}\hline \hline
Kick angle $(^\circ)$ & Bound mass & $|\tilde{L}_b|$ & $\vec{L}_b$ & Tilt angle $(^\circ)$ & Max. Luminosity\\ \tableline 
15 & 0.73 & 0.543 & (0.135,0.009,0.526) & 14.3 & 0.016\\
30 & 0.65 & 0.444 & (0.166, 0.011, 0.412) & 21.9 & 0.024 \\
45 & 0.60 & 0.359 & (0.146, 0.010, 0.327) & 24.2 & 0.039 \\
60 & 0.57 & 0.296 & (0.107, 0.006, 0.276) & 21.3 & 0.073 \\
None & 0.9999 &  0.830 & \tablenotemark{a} & \tablenotemark{a} & 0.015/0.004\tablenotemark{b} \\ \hline
\end{tabular}
\tablenotetext{a}{For the unkicked disk, $|\tilde{L}^x|,~|\tilde{L}^y| < 10^{-6}$, and thus $\theta_{tilt}<1.0e-6$ as well.}
\tablenotetext{b}{For the unkicked run, there is a brief burst of internal energy generation when the dynamical effects are turned on, yielding an internal energy generation rate $d\tilde{E}_{\rm INT}/d\hat{t}=0.015$, but thereafter the maximum steady state luminosity is $d\tilde{E}_{\rm INT}/d\hat{t}=0.004$.}
\end{center}
\end{table}

Once the initial disk is relaxed, it is allowed to evolve dynamically until $\hat{t}_{kick}=0.8$ before a kick is applied, except for a single unkicked control case we evolve to ensure that the physical effects we attribute to the kick are not merely an inevitable consequence of the dynamical evolution.  In the discussion that follows, we define the time since the kick $\hat{t}_*$ as
\begin{equation}
\hat{t}_*\equiv \hat{t}-\hat{t}_{kick}.\label{eq:tstardef}
\end{equation}

\begin{figure}
	\includegraphics[width=\columnwidth]{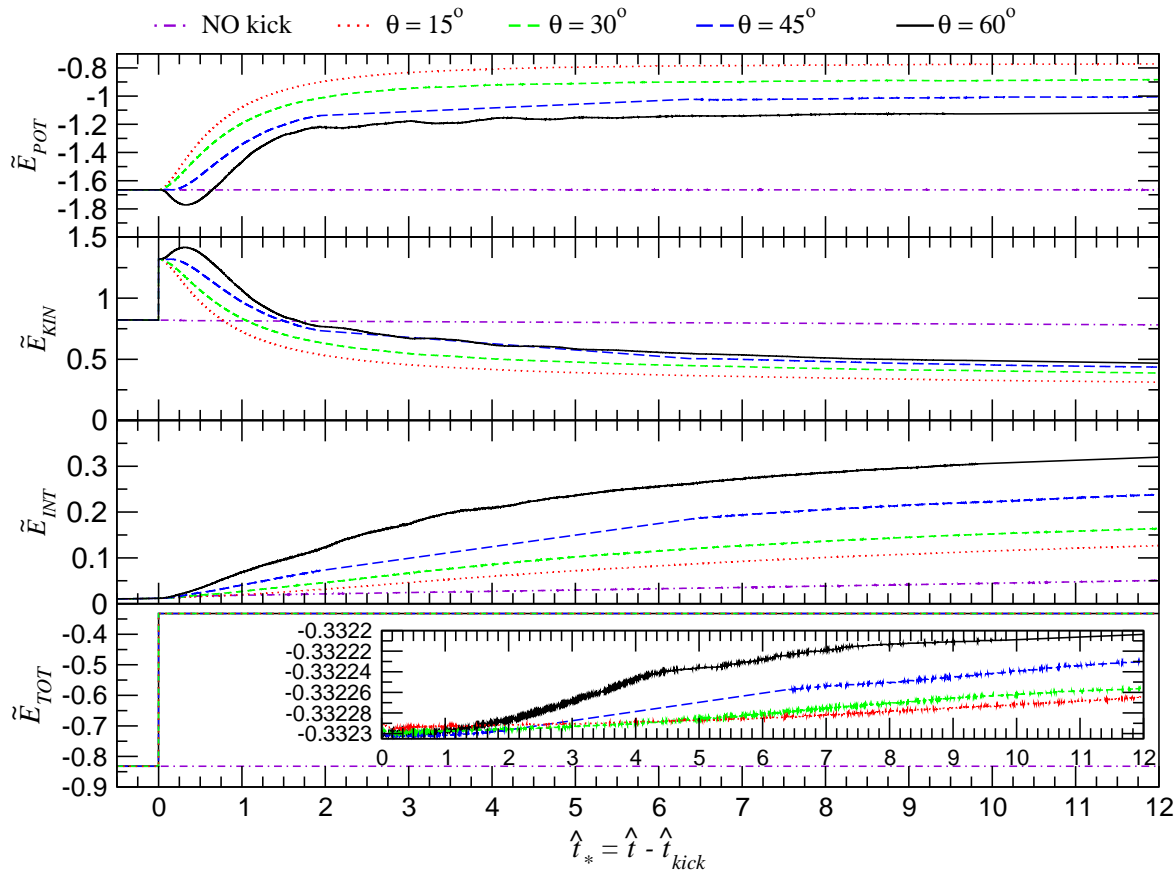}
        \caption{Evolution of the total potential energy (top panel), kinetic energy (second panel), internal energy (third panel), and total energy (bottom panel) for each of the runs.  The kick occurs at $\hat{t}_*=0$ (see Eq.~\protect\ref{eq:tstardef}). Each panel shows the unkicked control run (violet dotted-dashed line), and kicked cases of $15^\circ$(red dotted line), $30^\circ$(green dashed line), $45^\circ$(blue long-dashed line), and $60^\circ$(solid black line) away from vertical.  
        Energy preservation is almost perfect, with a total variation of no more than $\leq 0.03\%$ for any run, as shown in the inset plot in the bottom panel.}
        \label{fig:energies}
\end{figure}

As shown in the bottom panel of Fig.~\ref{fig:energies}, energy conservation is nearly exact for each of the runs, with total variation of no more than $0.03\%$ in the total energy after the kick in any of the runs.  Achieving this level of conservation is a consequence of two important components of the evolution scheme: the softened BH potential, described in Eq.~\ref{eq:cubicsplinepotential},  and the use of a Lagrangian-based variational scheme for evolving the smoothing length described in Sec.~\ref{sec:evoleqs}.  The former, which may be justified given the finite spatial extent of an SPH particle, prevents particles on highly eccentric orbits that approach very closely to the BH from attaining spurious energy during the periapse passage.  The latter, also used in \citet{Rossi:2009nk}, is required to allow for variable smoothing lengths without the energy varying on the same timescale as the smoothing lengths themselves.

As can be seen in Fig.~\ref{fig:energies}, the overall level of internal energy generation within the unkicked disk is approximately $30\%$  that of the most vertical kick configuration we consider ($15^\circ$ away from vertical), and roughly six times less that of the $60^\circ$ kick simulation.  We infer that while some of the heating we observe is an inevitable consequence of the disk evolution, the majority may be attributed to the kick and its aftermath, especially for cases where the kick is closer to the original disk plane.  Similarly, the changes in the kinetic and potential energy seem to be almost entirely a result of the kick.

\begin{figure}
	\includegraphics[width=\columnwidth]{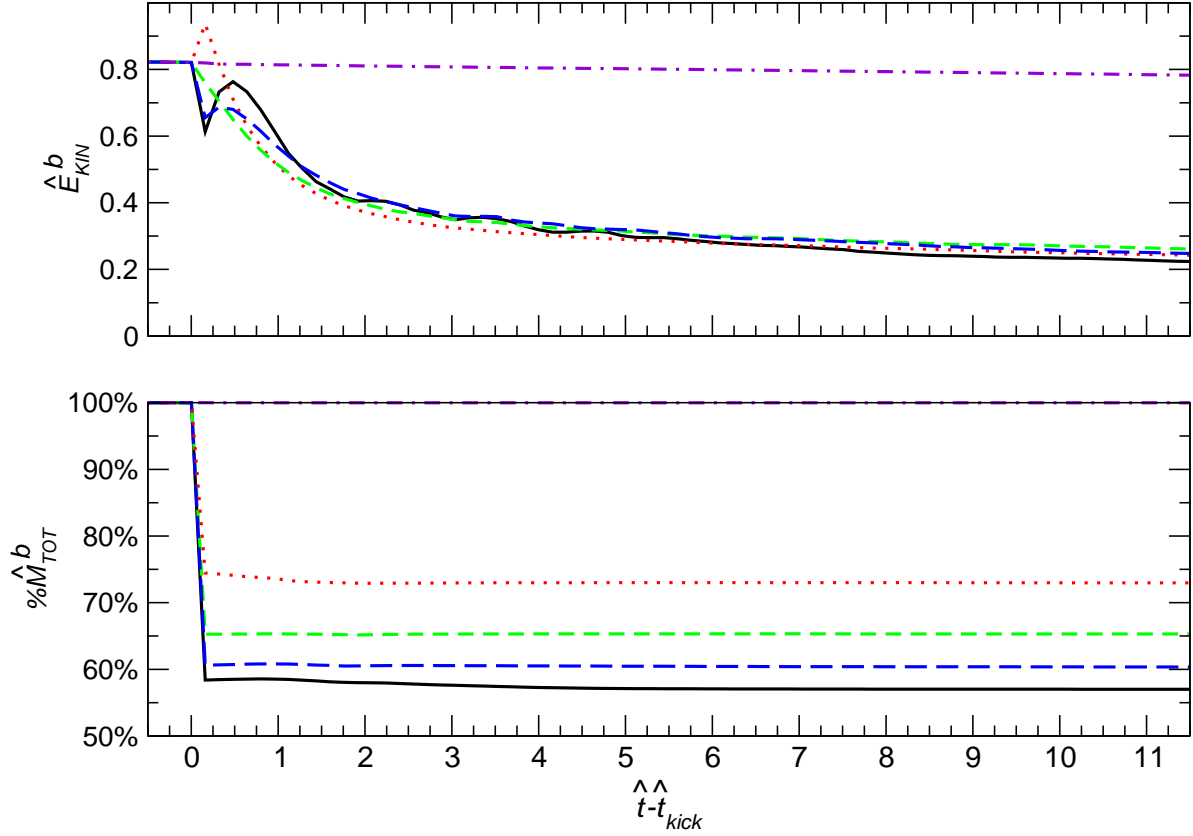}
        \caption{Evolution of the kinetic energy (top panel) and mass (bottom panel), relative to that of the initial disk, of the particles that remain bound to the SMBH.
       Conventions are as in Fig.~\protect\ref{fig:energies}, and the binding energy prescription is given by Eq.~\protect\ref{eq:bound}.}
        \label{fig:BndQts}
\end{figure}

By the end of our simulations, the kicked accretion disks have nearly reached a steady-state, as indicated by Fig.~\ref{fig:energies} (for global quantities) and Fig.~\ref{fig:BndQts} (for bound matter).  In general, the more oblique the kick, the more the resulting disk generates thermal energy, and, correspondingly, the deeper the potential energy well characterizing the disk.  While the {\em total} kinetic energy is nearly uniform among all the kick angles we consider, we note that the bound mass is smaller for more in-plane kicks given the initial configuration we chose, and thus the specific kinetic energy of the disk increases with the obliquity of the kick.

Fig.~\ref{fig:BndQts} shows the kinetic energy and relative mass of the bound portion of the disk after the kick for the different kick angles we considered as well as for the unkicked model. Clearly for the unkicked case, all the disk remains bound to the black hole, while for the different kick angles virtually all of the unbinding occurs at the moment of the kick.  The exact bound fractions are determined by our choice of initial disk configuration; our bound disk masses are particularly sensitive to the angle subtended by the bound region of the disk  at radii corresponding to the maximum surface density.
Note that our definition of binding is the criterion
\begin{equation}
E_{\rm POT,i}+E_{\rm KIN,i}+E_{\rm INT,i} = m_i\left(\Phi_i+\frac{|v_i|^2}{2}+\frac{3A_i\rho_i^{2/3}}{2}\right)<0\label{eq:bound},
\end{equation}
since the internal energy of an SPH particle on an otherwise bound trajectory would eventually lead to adiabatic expansion and unbinding of the constituent gas.  Disk heating does lead to some additional unbinding of material in most runs, ranging from no additional mass loss at all up to $1.5\%$ of the total mass, most of which occurs shortly after the kick.

Shortly after the kick, at times $0\le \hat{t}_*\lesssim 1$, the net change in the disk's kinetic and potential energies is strongly dependent on the kick angle, as can be seen in Fig.~\ref{fig:BndQts}.  This is a purely geometric result that can be described in terms of a simple collisionless model (see Sec.~\ref{sec:collisionless}): For a vertical kick, the entire mass of the disk is receding away from the BH immediately after the kick, and this remains nearly true for kicks near vertical.  For more oblique kicks, where the bound component of the disk is drawn primarily from the part of the disk whose rotational velocity is aligned with the kick itself, a substantial part of the mass finds itself on orbits approaching the BH immediately after the kick before collisions and shocks begin to circularize the new orbits.  Thus, for the oblique cases, we see an instantaneous decrease in the kinetic energy owing to the kick itself, followed by a rapid increase of larger magnitude as matter falls toward the BH.  For more vertical cases, the effect is reversed, and we see an instantaneous jump in kinetic energy followed by a gradual turnaround and decrease.  In both cases, the potential energy evolves accordingly, becoming more negative for the oblique cases relative to the vertical cases.

It is well-understood from previous calculations that the post-merger disk will be substantially tilted with respect to the initial equatorial plane, so we chose a simple prescription to define the post-kick disk plane.  Considering only the bound particles, as defined by Eq.~\ref{eq:bound}, we calculate the angular momentum of the bound component of the disk with respect to the black hole, yielding the results shown in Table~\ref{table:runs}.   Labeling this as the $z$-prime direction $\hat{z}'$, we define $\hat{x}'$ to be the original $x$-direction with the parallel projection of $\hat{z}'$ subtracted away, and the $\hat{y}'$-direction to be the cross product of the other two new coordinate directions:
\begin{eqnarray*}
\hat{z}' = \frac{\vec{L}_{bound}}{|\vec{L}_{bound}|};~~\hat{x}'=\frac{\hat{x}-(\hat{x}\cdot\hat{z}')\hat{z}'}{|\hat{x}-(\hat{x}\cdot\hat{z}')\hat{z}'|};~~\hat{y}'=\hat{z}'\times\hat{x}'.
\end{eqnarray*}
In all of the plots that follow, radii are assumed to represent cylindrical radii in the primed coordinate system.

\begin{figure}
\begin{center}
	\includegraphics[trim=0cm 10cm 0cm 0cm, clip,width=0.95\columnwidth]{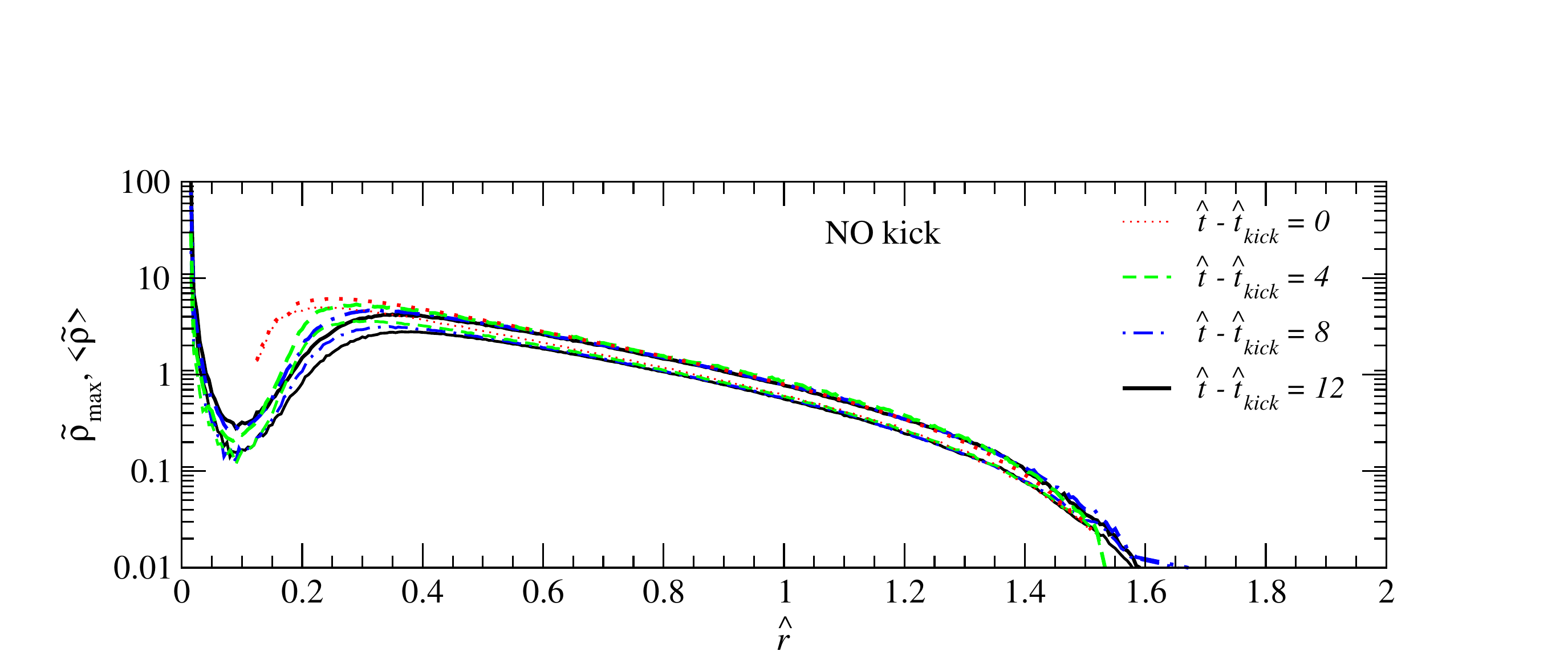}
	\vspace{-2cm}
	\includegraphics[trim=0cm -2cm 0cm 1.5cm, clip,width=0.95\columnwidth]{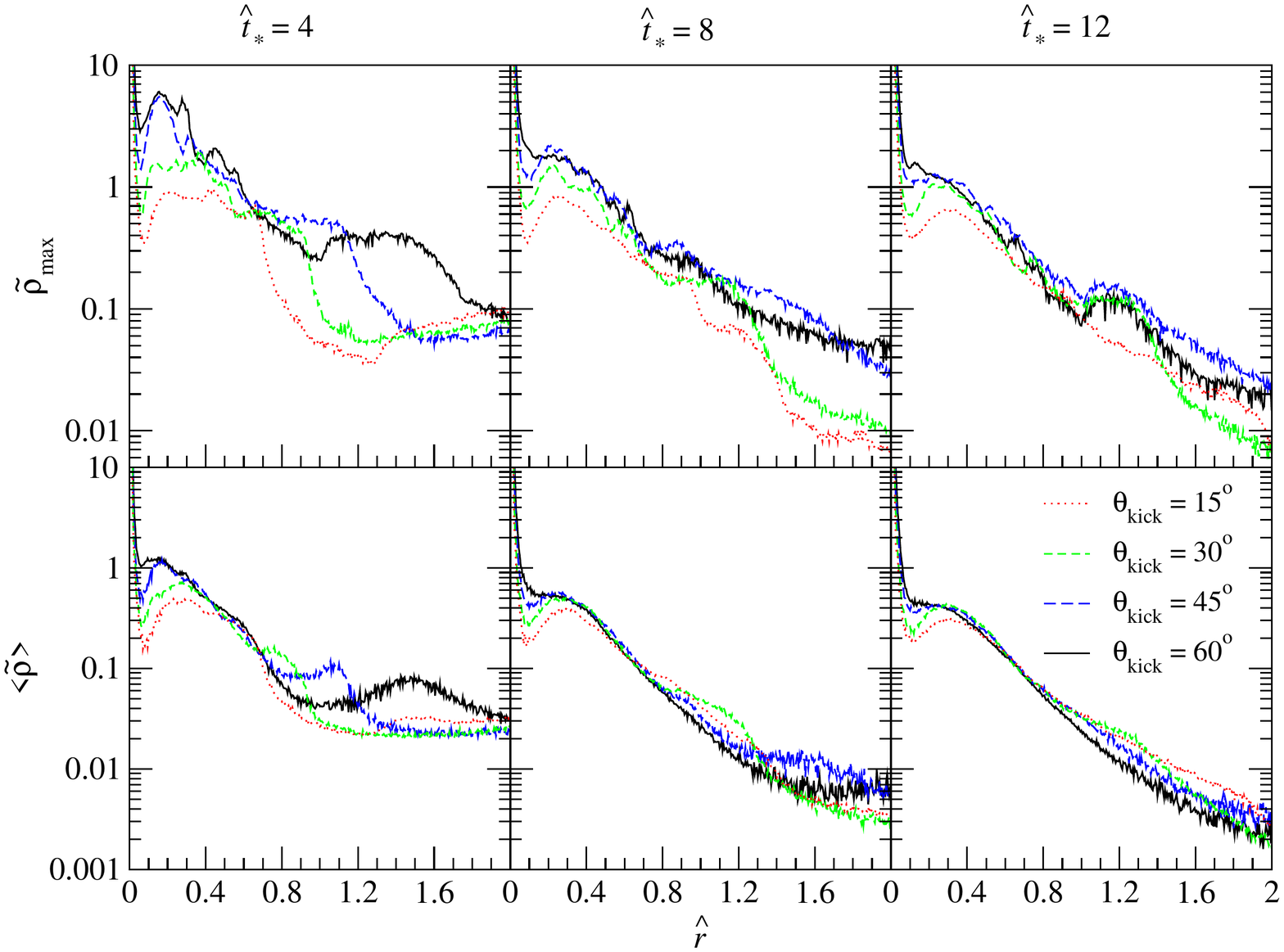}
\end{center}
        \caption{SPH particle-based density profile for the  unkicked run (top panel), showing the maximum (thick lines) and average SPH densities (thin lines) for particles binned with respect to cylindrical radius at $\hat{t}_*=0$ (red dotted lines), $\hat{t}_*=4$ (green dashed lines), $\hat{t}_*=8$ (blue dot-dashed lines), and $\hat{t}_*=12$ (black solid lines) .  Note that the $\hat{t}_*=0$ configuration presents the common pre-kick state for all of our runs.   There is a much more substantial density profile evolution in the kicked models (bottom panels).  From left to right we show the evolution of our kicked models at $\hat{t}_*=4$ (left), $\hat{t}_*=8$ (center), and $\hat{t}_*=12$ (right), with the maximum SPH density shown in the upper panels and the average SPH density in the lower panels.  The curves follow the conventions of Fig.~\protect\ref{fig:energies}. }
        \label{fig:logRhoI-rpos}
\end{figure}

\begin{figure}
\begin{center}
	\includegraphics[trim=0cm 1cm 0cm 1.5cm, clip=true, width=0.95\columnwidth]{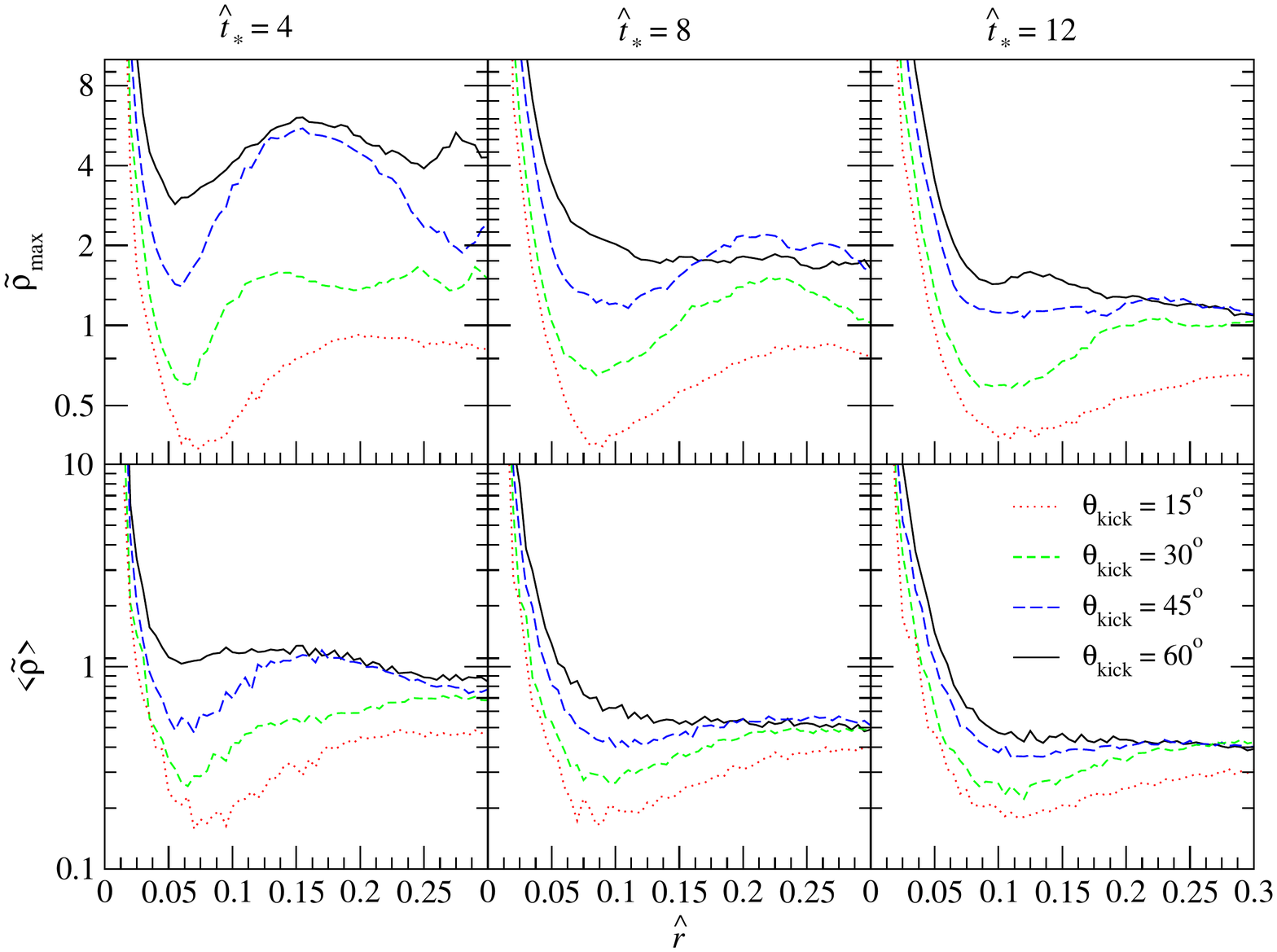}
\end{center}
        \caption{SPH particle-based densities for the innermost part of the accretion disks.  All conventions follow those in Fig.~\protect\ref{fig:energies} and the bottom panel of Fig.~\protect\ref{fig:logRhoI-rpos}.}
        \label{fig:logRhoI-rpos2}
\end{figure}

To confirm the validity of the ``gap-filling'' model we discussed in Sec.~\ref{sec:collisionless}, we show the SPH densities of the particles in the inner disk as a function of the cylindrical radius in Figs.~\ref{fig:logRhoI-rpos} and \ref{fig:logRhoI-rpos2}.  Turning first to the unkicked control model in the top panel of the former, we see that there is relatively little density evolution except at the innermost edge of the disk, where viscous dissipation of angular momentum leads to an accretion of particles toward the SMBH.  A density peak does form at the center, but with smaller densities  than the kicked runs at any given radius $\hat{r}\lesssim 0.1$ during the duration of our simulations.

Considering the kicked runs next, we observe that the filling of the gap proceeds as predicted above.  A wave of particles at relatively high densities ($\tilde{\rho}\sim 1-10$) is observed moving inward at $\hat{t}_*=4$ for the two most oblique kick angles we consider, namely $\theta=45^\circ$ and $60^\circ$, particularly the latter.  No such densities are ever found except in the immediate vicinity of the BH ($\hat{r}\lesssim 0.05$) for the more vertical kicks ($\theta=15^\circ$ and $30^\circ$).   Part of this effect  is a simple matter of the larger post-kick periastron radii in the more vertical cases (see Fig.~\ref{fig:rperi}).     Also, since the entire inner region of the disk remains bound in these cases, it forms an ``inner ring'' that will block any infalling matter  from reaching radii $\hat{r}\lesssim 0.1-0.2$.  For the more oblique cases, not only does some matter accrete promptly, but a gap is formed in the inner region corresponding to the initially retrograde portion of the disk (i.e., the material initially located near $\phi=-90^\circ$) that allows material to be channeled more easily toward  radii  $\hat{r}\lesssim 0.1$.  This resulting density enhancement surrounding the BH is quite persistent; at $\hat{t}_*=12$, which represents several orbital timescales for the innermost edge of the disk, there is a factor of five difference in the density at $\hat{r}\le 0.1$  between the $60^\circ$ and $15^\circ$ runs.  We note that there is some ``crystallization'' and pattern formation of the particles located closest to the BH, which is inevitable when a small number of SPH particles are located near the edge of a density distribution, but that we still are able to resolve a smooth density field there.  

For the most oblique model, there is a clear oscillation in the maximum density at early times, which is highlighted in Fig.~\ref{fig:logRhoI-rpos2}, showing the densities for the innermost region of the accretion disk.  These features are an indication of the spiral density waves present in the disk and leave a clear imprint on the resulting thermodynamic evolution and emission properties we predict, as we discuss in more detail below. The general behavior of these density pulses, which move inward with time, is to increase the central density of the disk, as we can see by the closing of the gap between the densities for the $45^\circ$ and $60^\circ$ runs between $\hat{t}_*=8$ and $\hat{t}_*=12$.    For the more vertically kicked runs, where spiral waves are much weaker, there is very little sign of rapid accretion of material to the center.

In setting up the runs shown here, it became evident that the BH softening potential we apply, Eq.~\ref{eq:cubicsplinepotential}, can play an important role in suppressing spurious energy fluctuations.  Indeed, in simulations without a BH softening potential, we find that the innermost particles around the BH clump together, similarly to the so-called pairing instability of SPH \citep{Price:2010hv} but with more than two particles per clump.  These clumps 
form quasi-stationary ``bubbles,'' where mutual pressure interactions keep any particle from approaching within about a smoothing length of the black hole.  This behavior is robust against several different choices of the SPH smoothing kernel definition and evolution schemes for the smoothing length in time.  When outside interactions finally ``pop'' the bubble, and other particles are able to flow inward to smaller radii, the measured kinetic and potential energies are seen to jump by substantial amounts because of a handful of particles, even though the total energy remained well conserved.

\begin{figure}
\begin{center}
	\includegraphics[trim=0cm 10cm 0cm 0cm, clip, width=\columnwidth]{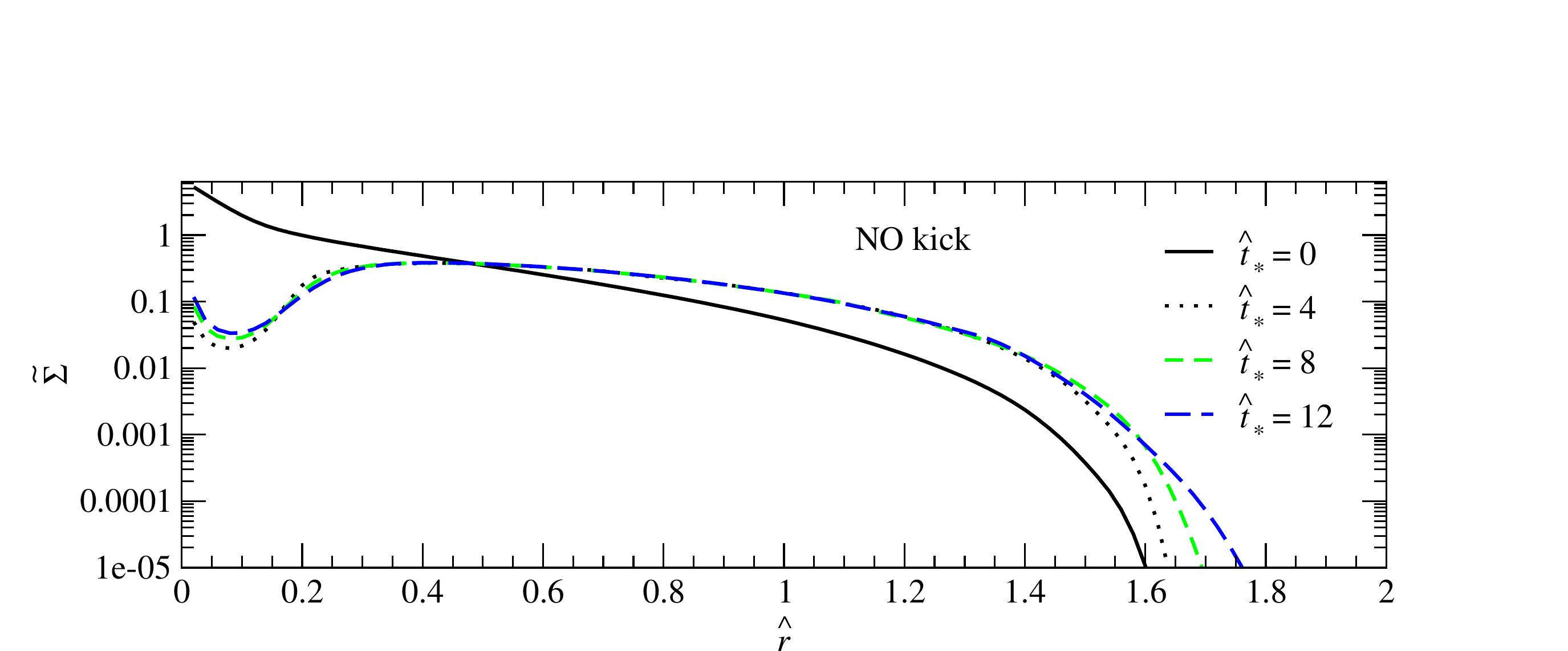}
        \vspace{-2cm}
	\includegraphics[trim=0cm -1cm 0cm 2cm, clip, width=\columnwidth]{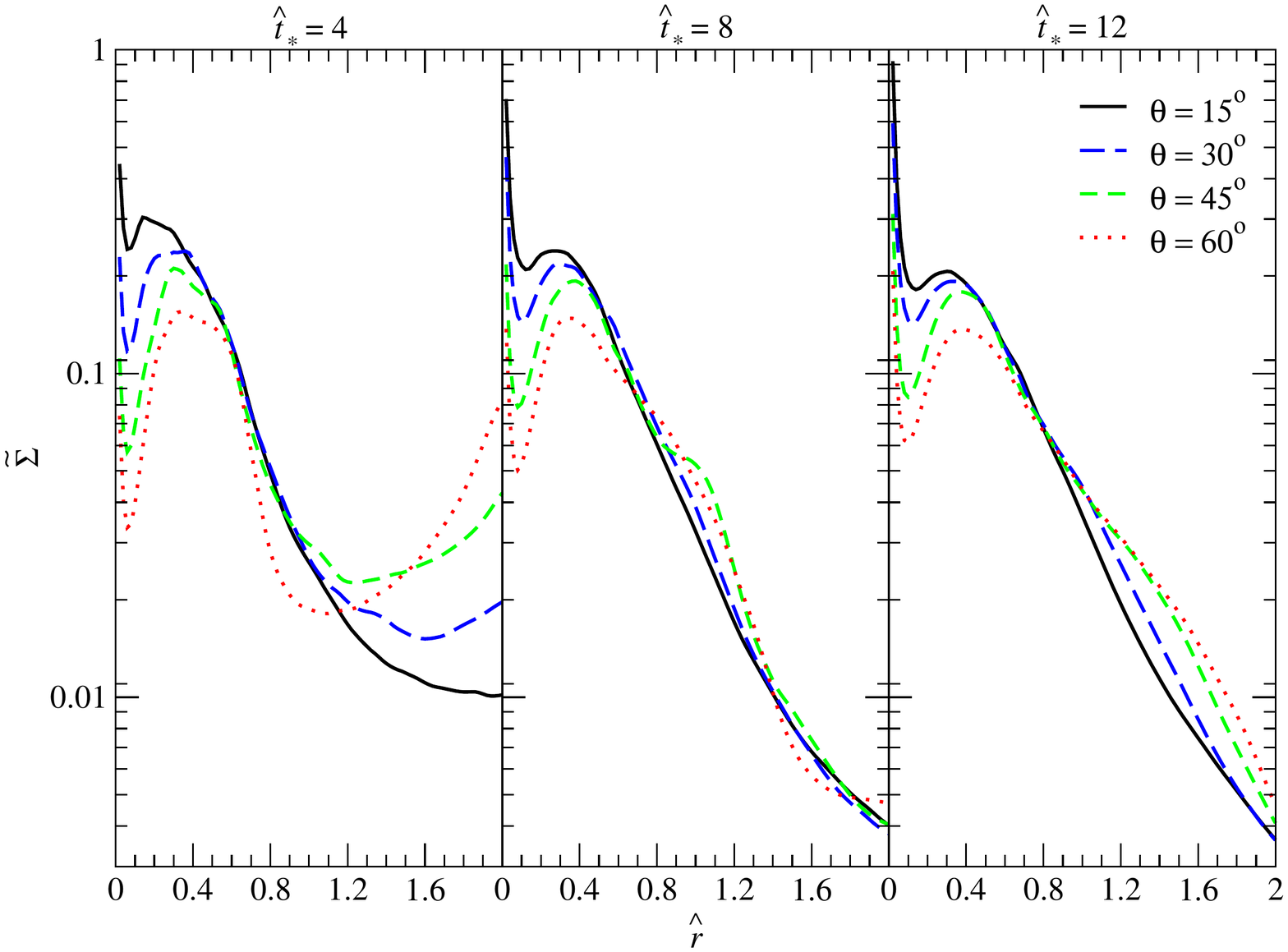}
\end{center}
        \caption{Radial profiles (averaged over azimuthal angle) of the surface densities (in log scale) for the initial disk model and the no-kick model (top panel) and for kick angles of 15, 30, 45 and 60 degrees; at $\hat{t}_*=4, 8, 12$ (lower panels).}
        \label{fig:surfDens-rad_profiles}
\end{figure}

For a slightly different view of the accretion disks, we show the azimuthally averaged radial surface  density  profiles $\tilde{\Sigma}(\hat{r})$ in Fig.~\ref{fig:surfDens-rad_profiles}.  We see that the surface densities are markedly different in the inner region, with the oblique kicks leading to persistently higher surface densities by at least a factor of five compared to more vertical kicks for $\hat{r}\lesssim 0.2$ throughout the course of the simulation.  In the outer regions of the bound disk, the surface density trend is reversed but much less dramatic:  the more vertical kicks have slightly larger surface densities at a given radius than the oblique kicks, but the variation is never more than a factor of two once the disk begins to relax again at $\hat{t}_*\gtrsim 4$.  The ratio of the initial surface density to the postkick surface density is relatively constant over a wide range of radii, from $0.4\lesssim \hat{r}\lesssim 1.5$, but the postkick disk extends much further, since the same angular momentum exchange processes that funnel matter toward the BH at the inner edge of the disk also help to circularize it to larger radii at the outer edge of the bound region.

Our results also allow us to make some rough conclusions about the opacities of our disks, though we note we do not include any radiation transport effects nor radiative cooling in the simulations of Sec.~\ref{subsec:heat}.  Because the disks are hot and diffuse throughout, we expect the Thomson opacity for an ionized plasma to be a reasonable approximation.  Whether or not the disk is optically thick ($\tau>>1$) depends not only on the dimensionless surface density $\tilde{\Sigma}$ but also on the values of the disk mass $m_{\rm disk}$, the BH mass $M_{\rm BH}$, and the kick velocity $v_{\rm kick}$ (see Eq.~\ref{eq:depth}).  For our reference model ($m_{\rm disk}=10^4M_\odot$, $M_{\rm BH}=10^8M_\odot$, and $v_{\rm kick}=10^8$cm/s), the optical depth is slightly larger than unity throughout the pre-kick disk and slightly below throughout the post-kick disk, in which $\Sigma_{post}(r)/\Sigma_{pre}(r)\approx 0.2-0.4$ for most of the area of the disk, $0.4\lesssim \hat{r}\lesssim 1.5$.  If the initial disk had a substantially smaller surface density, our model would predict that the post-kick disk would remain so as well, except in the very central region near the BH.  Meanwhile, an optically thick initial disk should produce a slightly less thick disk after the kick, extending outward to nearly the edge of the bound component.

\begin{figure}
\begin{center}
\begin{tabular}{ccccl}
& &
         $\hat{t}_* = 4$
        &
         $\hat{t}_* = 8$
        &
        \hspace{0.1\columnwidth}  $\hat{t}_* = 12$
\\ \raisebox{1.5cm}{No Kick} & $\hat{y}$ &
       \raisebox{-2cm}{\includegraphics[trim=7.8cm 4.5cm 7cm 4.5cm, clip=true, width=0.25\columnwidth]{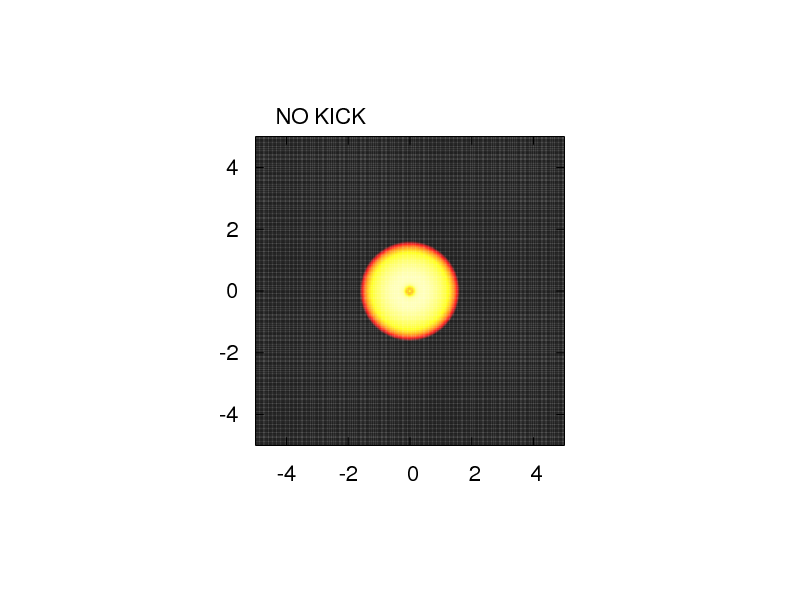}}
        &
        \raisebox{-2cm}{\includegraphics[trim=7.8cm 4.5cm 7cm 4.5cm, clip=true, width=0.25\columnwidth]{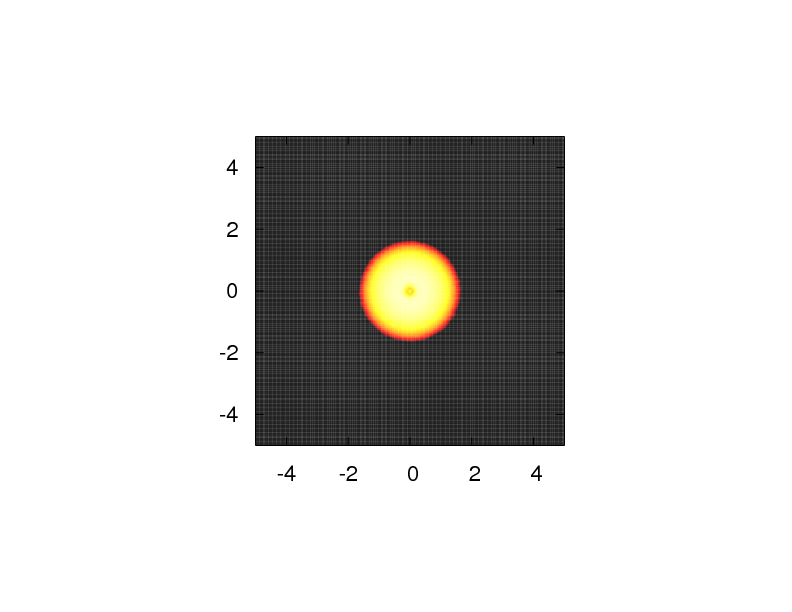}}
        &
        \raisebox{-2cm}{\includegraphics[trim=7.8cm 4.5cm 5cm 4.5cm, clip=true, width=0.29\columnwidth]{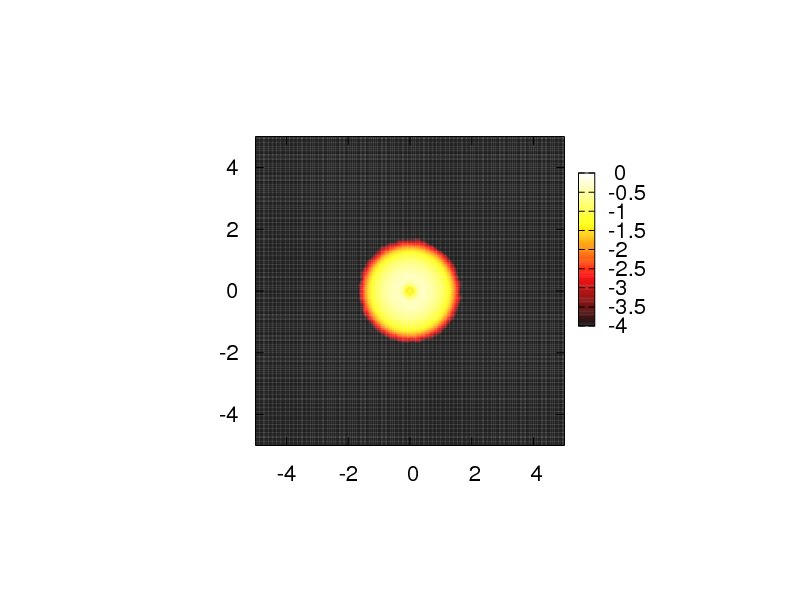}}
\\
\raisebox{1.5cm}{$\theta=15^\circ$} & $\hat{y}$  &
        \raisebox{-2cm}{\includegraphics[trim=7.8cm 4.5cm 7cm 4.5cm, clip=true, width=0.25\columnwidth]{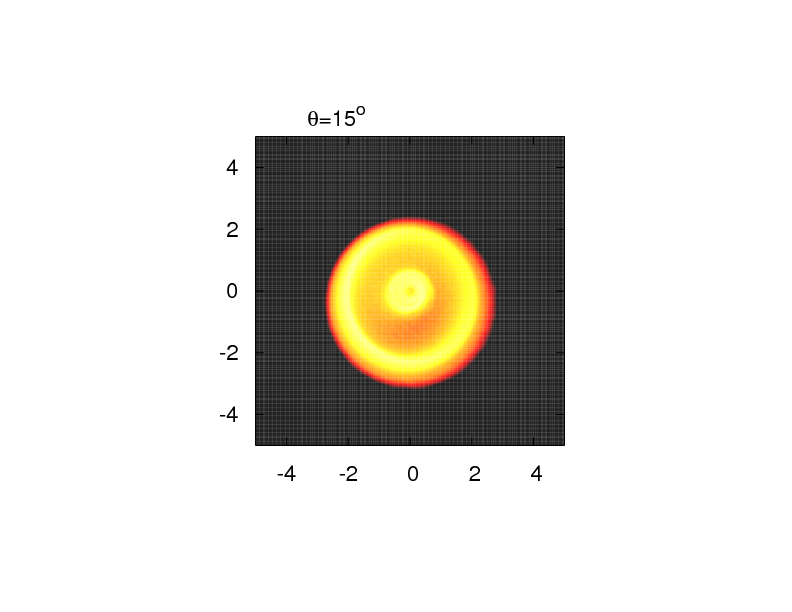}}
        &
        \raisebox{-2cm}{\includegraphics[trim=7.8cm 4.5cm 7cm 4.5cm, clip=true, width=0.25\columnwidth]{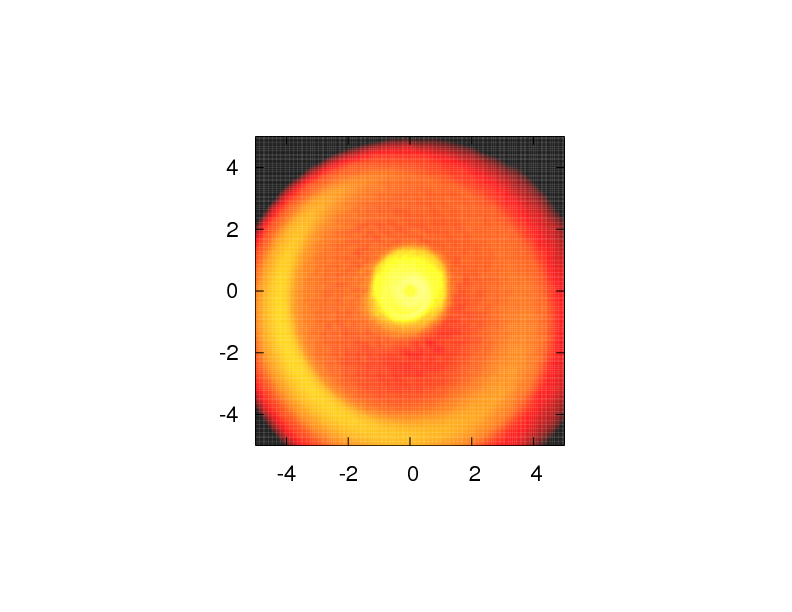}}
        &
       \raisebox{-2cm}{ \includegraphics[trim=7.8cm 4.5cm 7cm 4.5cm, clip=true, width=0.25\columnwidth]{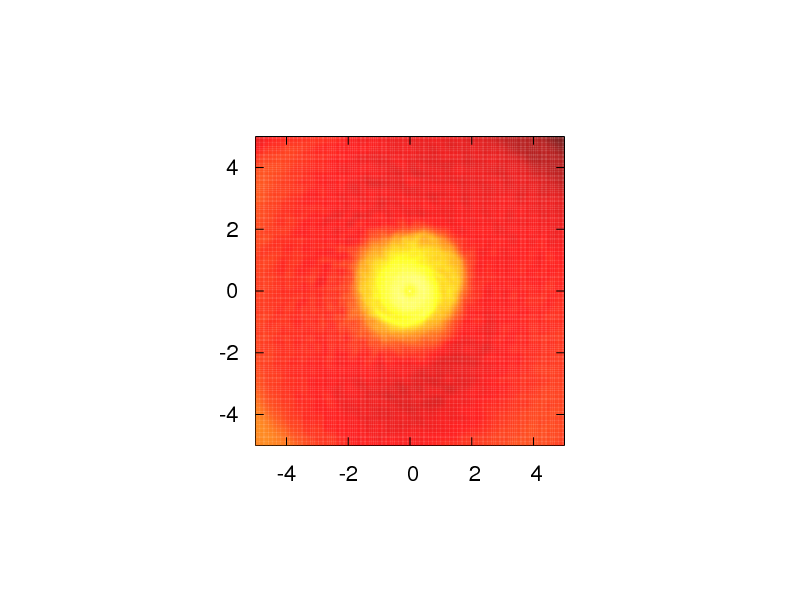}}
\\ \raisebox{1.5cm}{$\theta=30^\circ$} & $\hat{y}$&
        \raisebox{-2cm}{ \includegraphics[trim=7.8cm 4.5cm 7cm 4.5cm, clip=true, width=0.25\columnwidth]{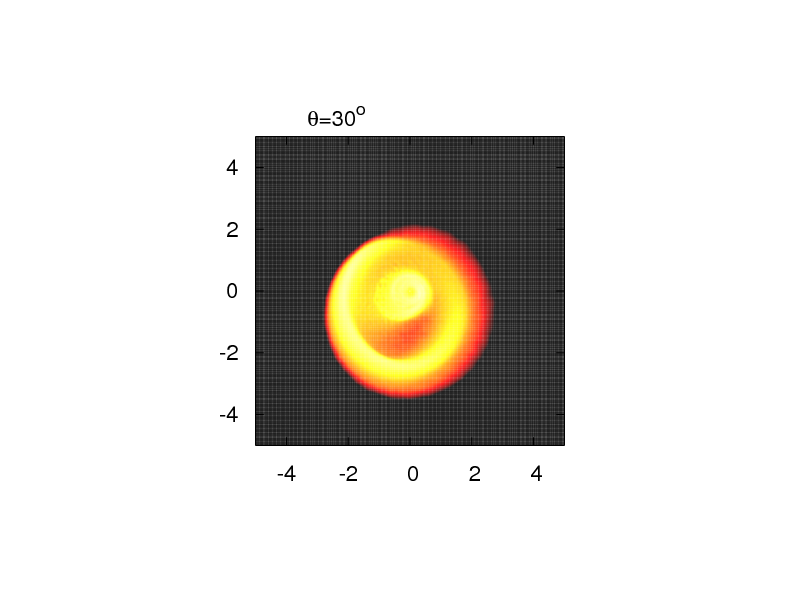}}
        &
        \raisebox{-2cm}{ \includegraphics[trim=7.8cm 4.5cm 7cm 4.5cm, clip=true, width=0.25\columnwidth]{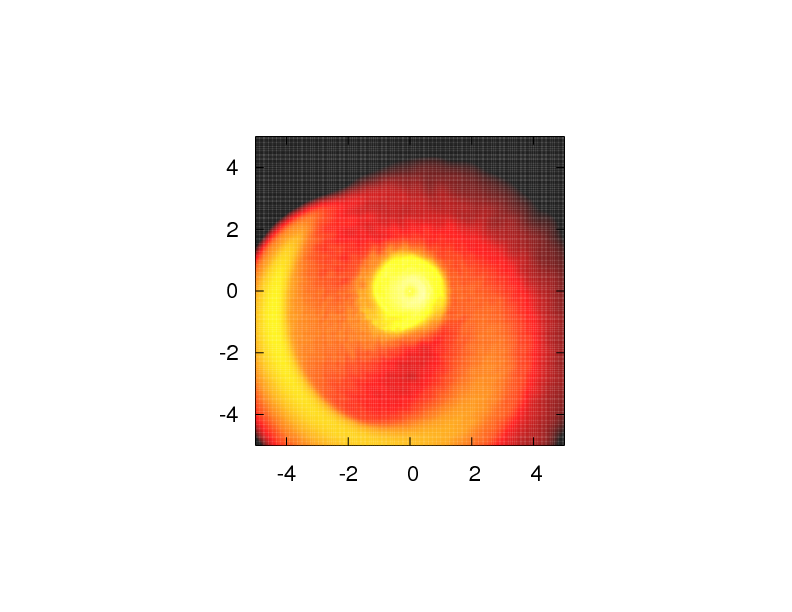}}
        &
        \raisebox{-2cm}{ \includegraphics[trim=7.8cm 4.5cm 7cm 4.5cm, clip=true, width=0.25\columnwidth]{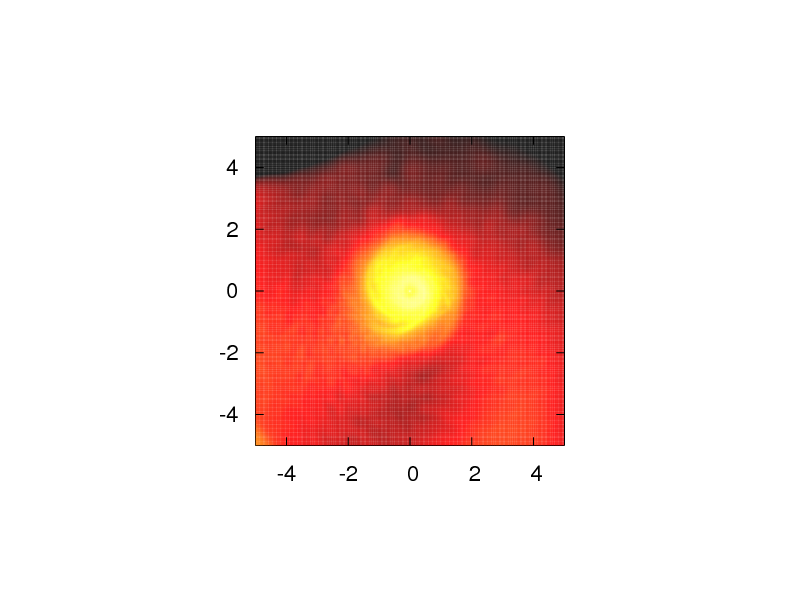}}
\\ \raisebox{1.5cm}{$\theta=45^\circ$} & $\hat{y}$&
        \raisebox{-2cm}{ \includegraphics[trim=7.8cm 4.5cm 7cm 4.5cm, clip=true, width=0.25\columnwidth]{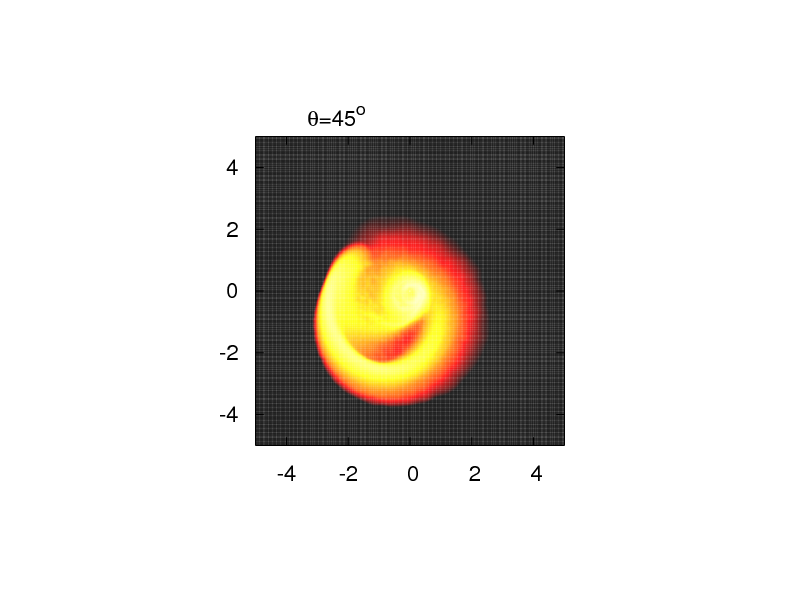}}
        &
       \raisebox{-2cm}{  \includegraphics[trim=7.8cm 4.5cm 7cm 4.5cm, clip=true, width=0.25\columnwidth]{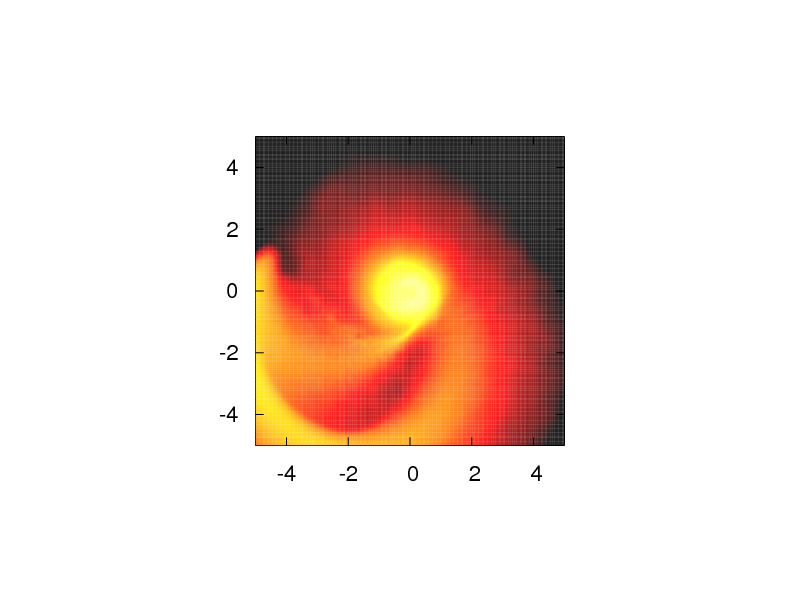}}
        &
        \raisebox{-2cm}{ \includegraphics[trim=7.8cm 4.5cm 7cm 4.5cm, clip=true, width=0.25\columnwidth]{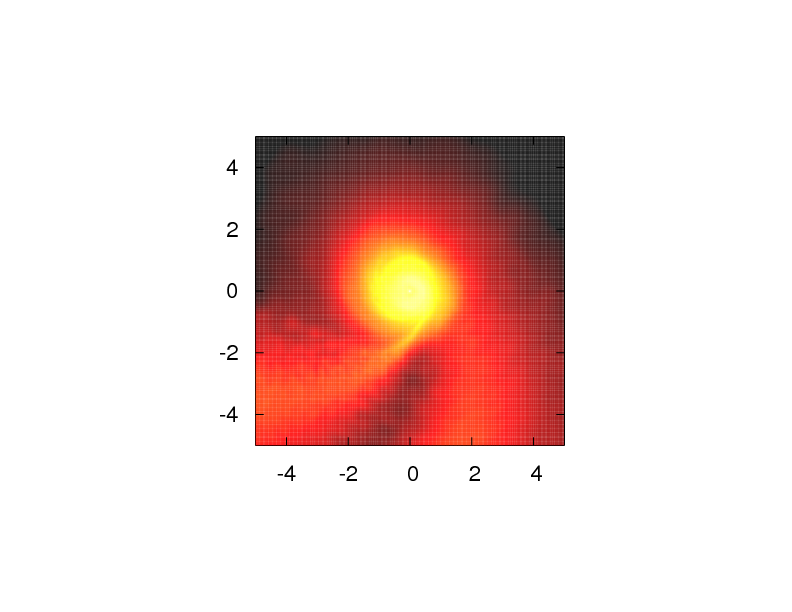}}
\\ \raisebox{1.5cm}{$\theta=60^\circ$} & $\hat{y}$&
       \raisebox{-2cm}{  \includegraphics[trim=7.8cm 4.5cm 7cm 4.5cm, clip=true, width=0.25\columnwidth]{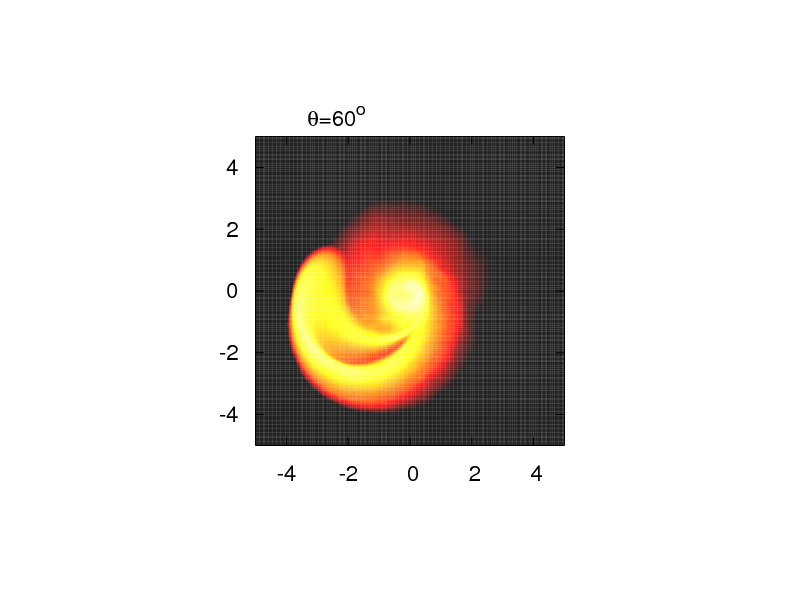}}
        &
       \raisebox{-2cm}{  \includegraphics[trim=7.8cm 4.5cm 7cm 4.5cm, clip=true, width=0.25\columnwidth]{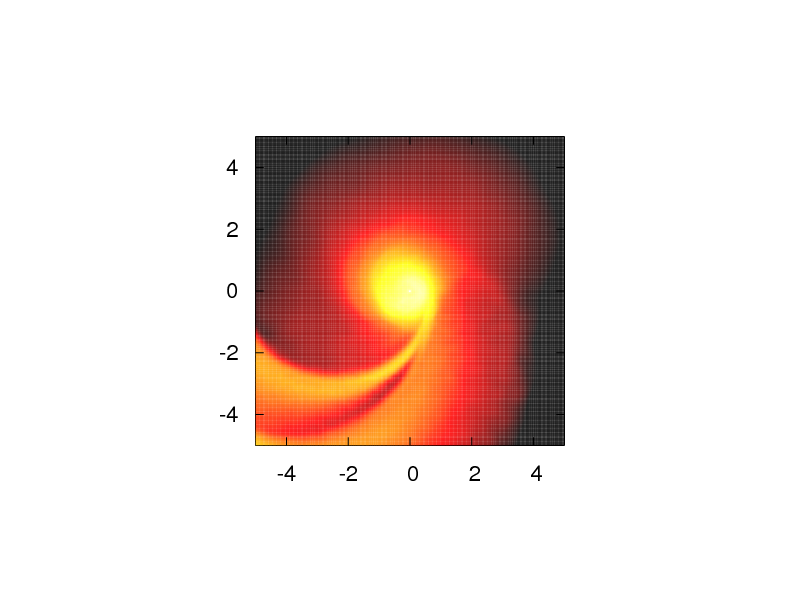}}
        &
        \raisebox{-2cm}{ \includegraphics[trim=7.8cm 4.5cm 7cm 4.5cm, clip=true, width=0.25\columnwidth]{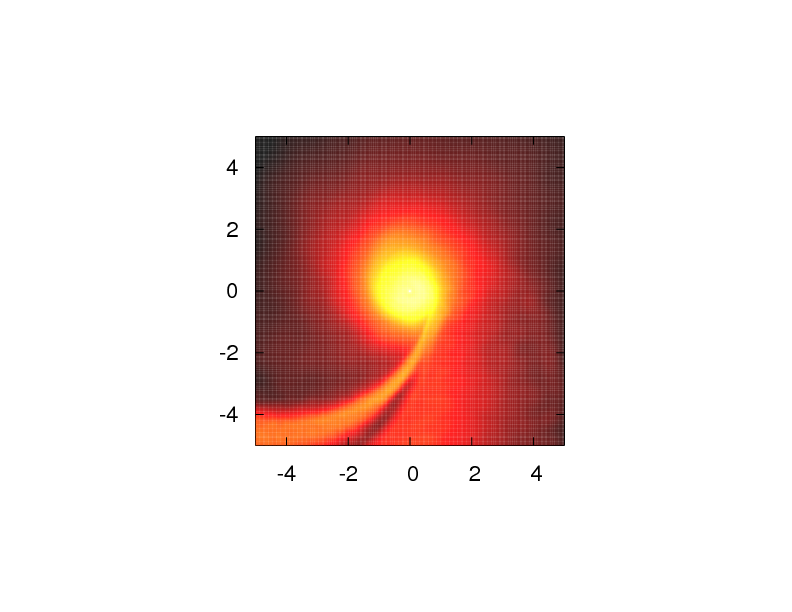}}
        \\& & $\hat{x}$ & $\hat{x}$ & \hspace{2cm}$\hat{x}$ \\
        
\end{tabular}
        \caption{Projected surface densities (in log scale) for No-kick, and kick angles of 15, 30, 45 and 60 degrees (rows from top to bottom respectively); at $\hat{t}_*=4, 8, 12$.         \label{fig:surfDens-map}
}
        \end{center}
 \end{figure}

\begin{figure}   
\begin{center}
\begin{tabular}{cccl}
 &
         $\hat{t}_* = 4$
        &
          $\hat{t}_* = 8$
        &
        \hspace{0.1\columnwidth}  $\hat{t}_* = 12$
\\  $\hat{y}$&
        \raisebox{-2cm}{ \includegraphics[trim=7cm 4.3cm 7cm 4.3cm, clip=true, width=0.25\columnwidth]{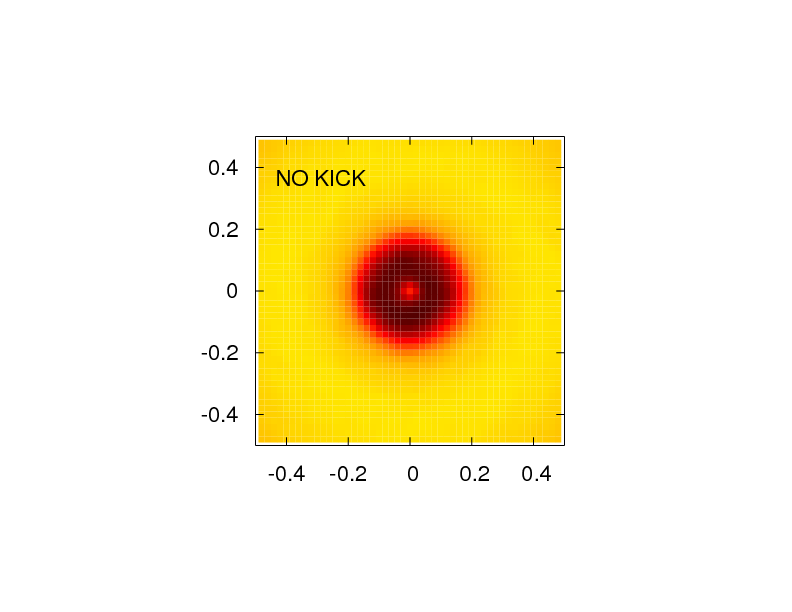}}
        &
        \raisebox{-2cm}{ \includegraphics[trim=7cm 4.3cm 7cm 4.3cm, clip=true, width=0.25\columnwidth]{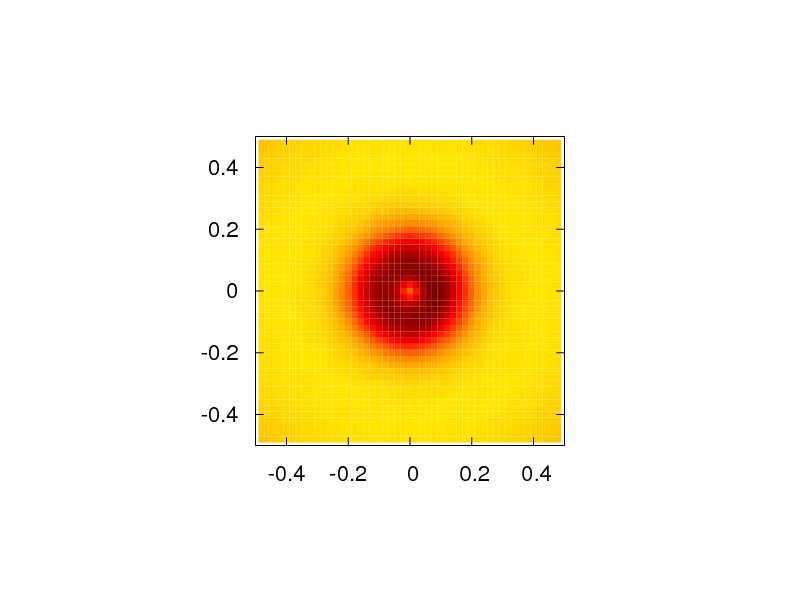}}
        &
        \raisebox{-2cm}{ \includegraphics[trim=7cm 4.3cm 5cm 4.3cm, clip=true, width=0.29\columnwidth]{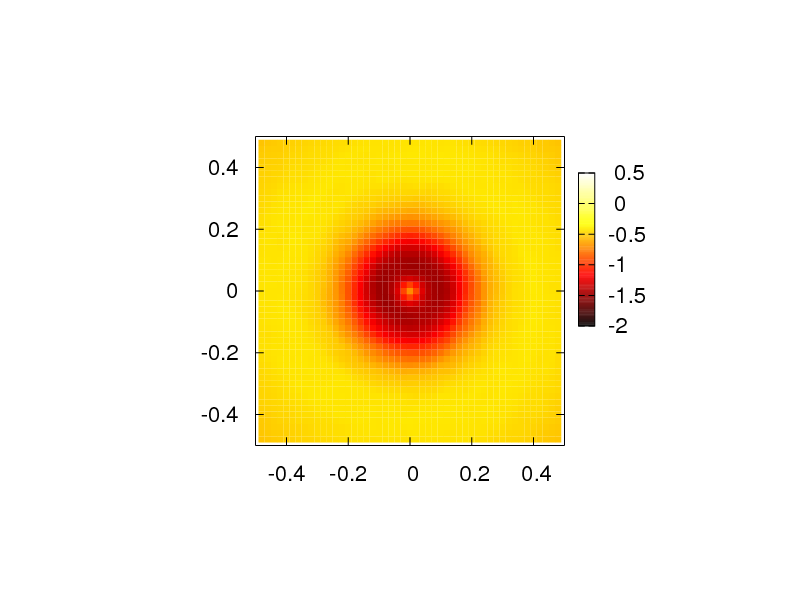}}
\\  $\hat{y}$&
        \raisebox{-2cm}{ \includegraphics[trim=7cm 4.3cm 7cm 4.3cm, clip=true, width=0.25\columnwidth]{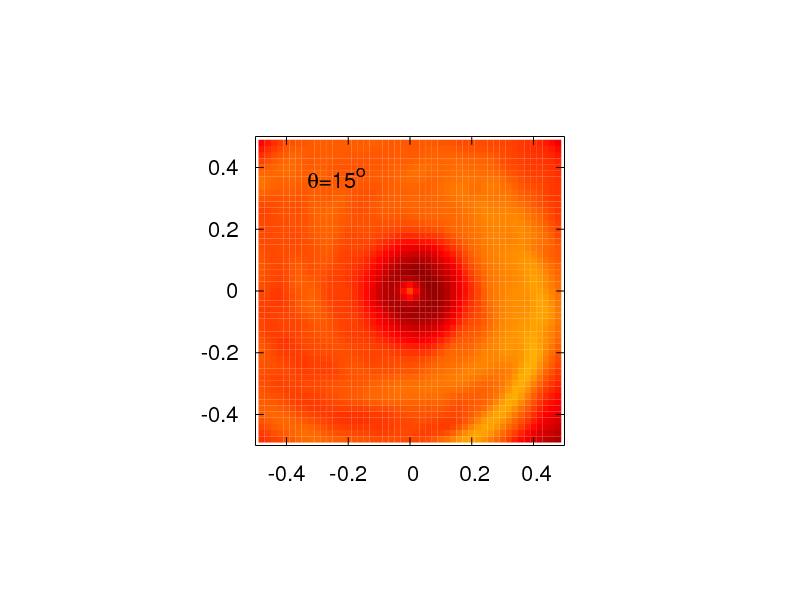}}
        &
        \raisebox{-2cm}{ \includegraphics[trim=7cm 4.3cm 7cm 4.3cm, clip=true, width=0.25\columnwidth]{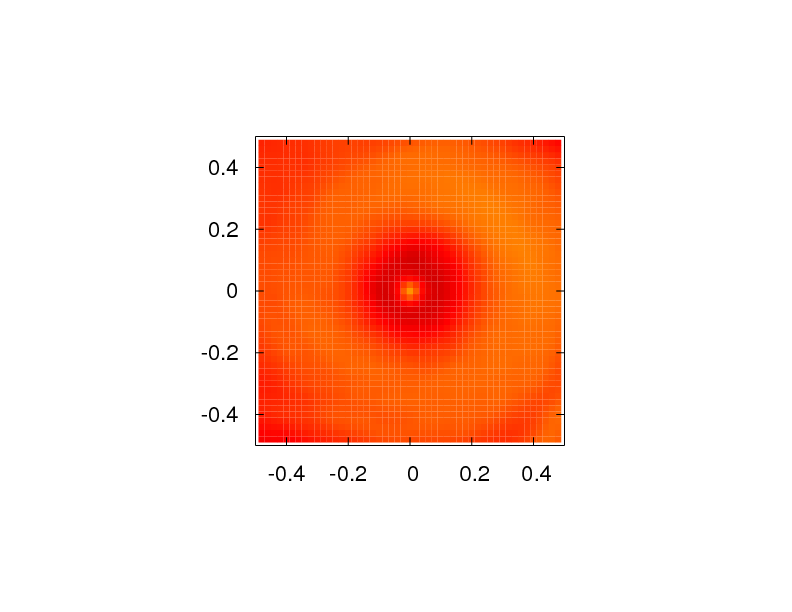}}
        &
        \raisebox{-2cm}{ \includegraphics[trim=7cm 4.3cm 7cm 4.3cm, clip=true, width=0.25\columnwidth]{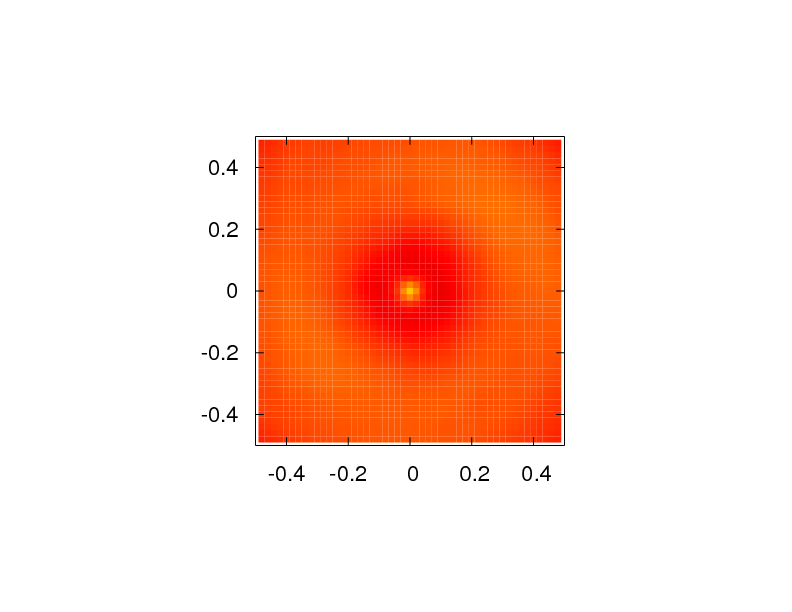}}
\\  $\hat{y}$&
       \raisebox{-2cm}{  \includegraphics[trim=7cm 4.3cm 7cm 4.3cm, clip=true, width=0.25\columnwidth]{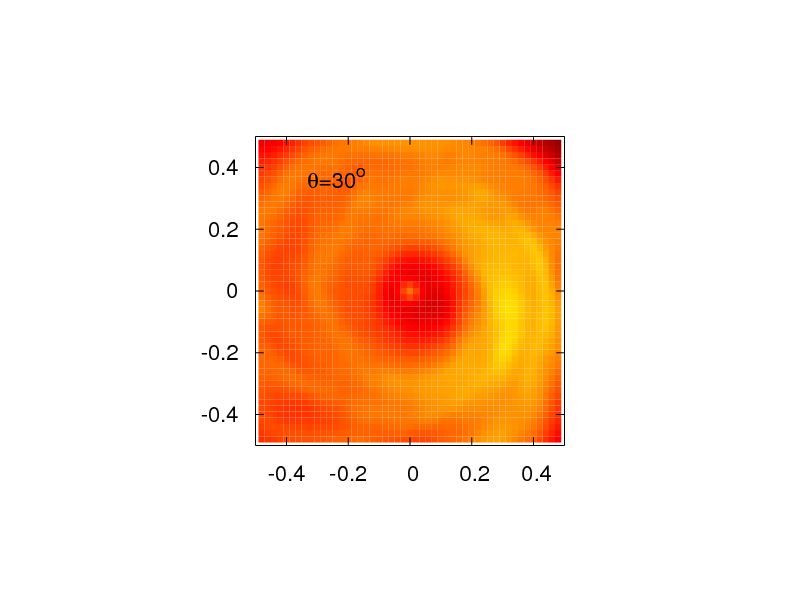}}
        &
        \raisebox{-2cm}{ \includegraphics[trim=7cm 4.3cm 7cm 4.3cm, clip=true, width=0.25\columnwidth]{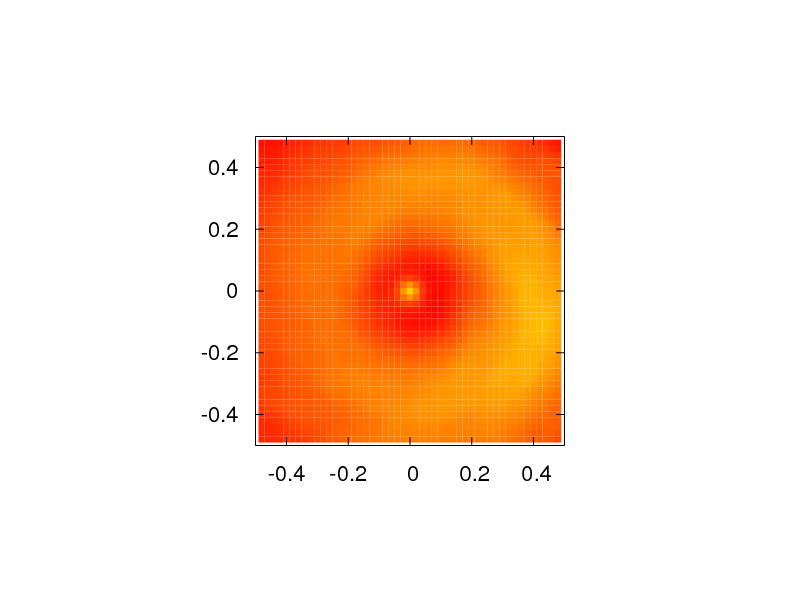}}
        &
        \raisebox{-2cm}{ \includegraphics[trim=7cm 4.3cm 7cm 4.3cm, clip=true, width=0.25\columnwidth]{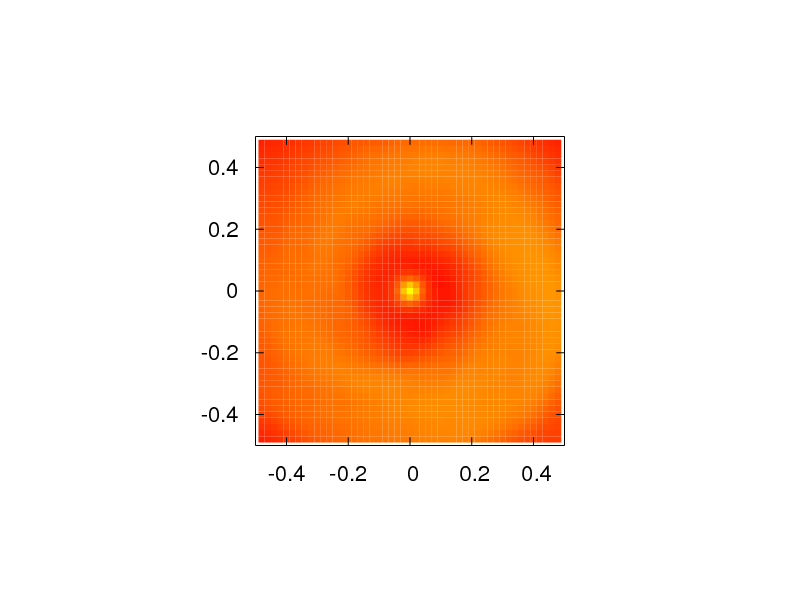}}
\\ $\hat{y}$&
        \raisebox{-2cm}{ \includegraphics[trim=7cm 4.3cm 7cm 4.3cm, clip=true, width=0.25\columnwidth]{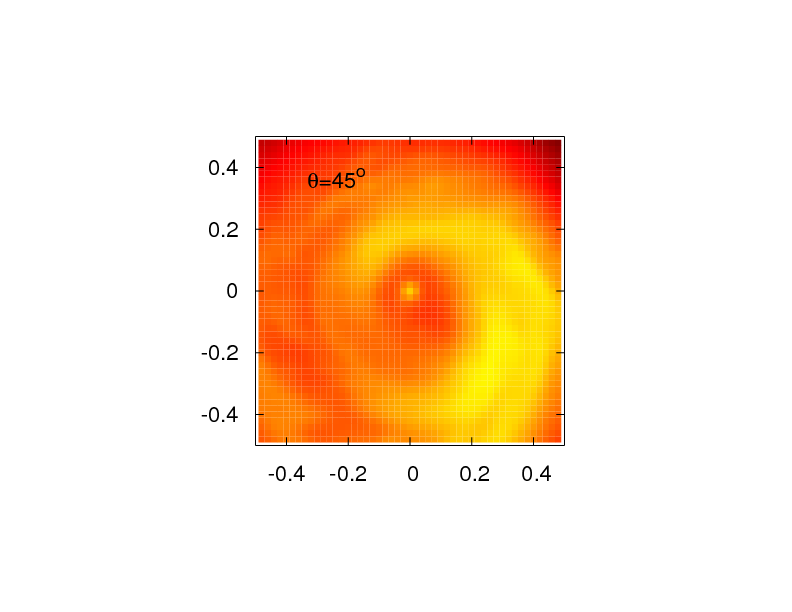}}
        &
       \raisebox{-2cm}{  \includegraphics[trim=7cm 4.3cm 7cm 4.3cm, clip=true, width=0.25\columnwidth]{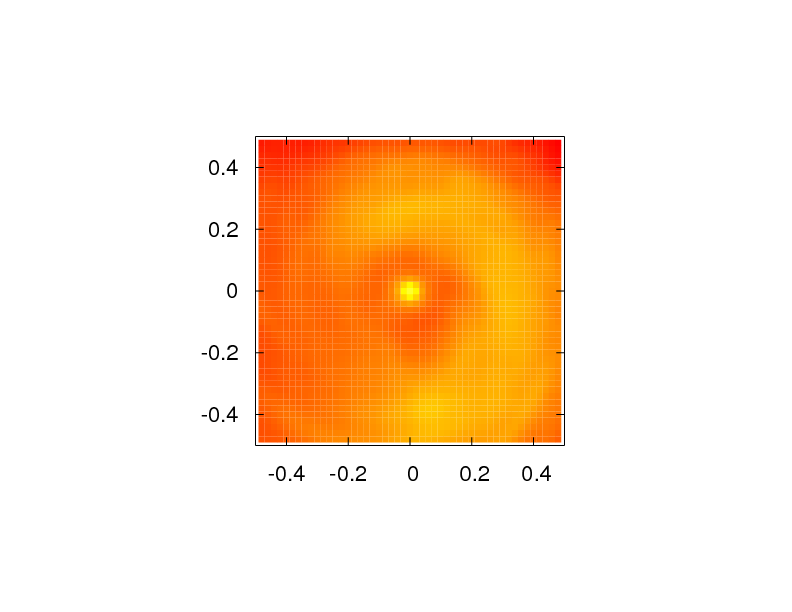}}
        &
       \raisebox{-2cm}{  \includegraphics[trim=7cm 4.3cm 7cm 4.3cm, clip=true, width=0.25\columnwidth]{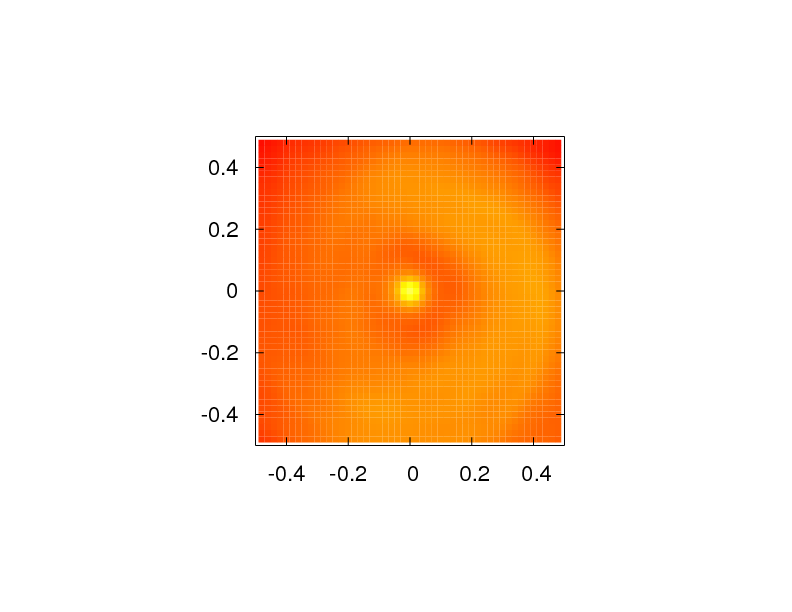}}
\\  $\hat{y}$&
       \raisebox{-2cm}{  \includegraphics[trim=7cm 4.3cm 7cm 4.3cm, clip=true, width=0.25\columnwidth]{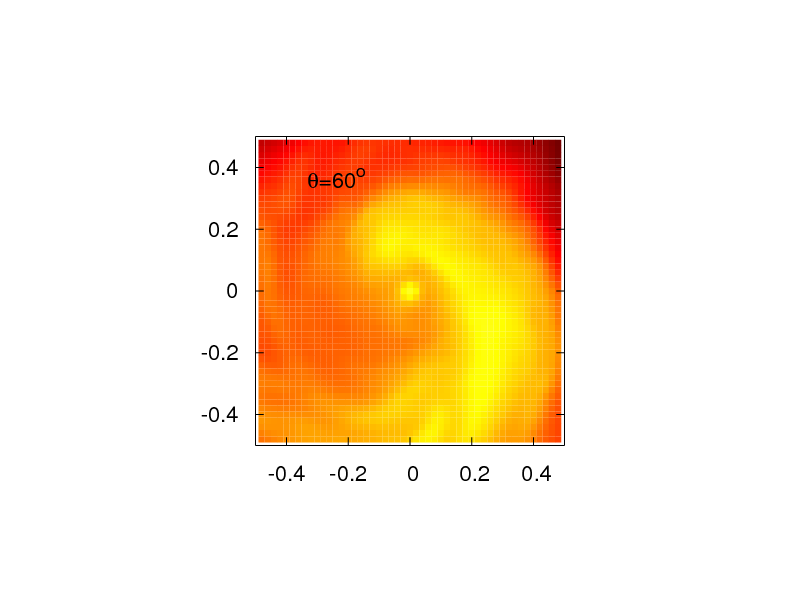}}
        &
       \raisebox{-2cm}{  \includegraphics[trim=7cm 4.3cm 7cm 4.3cm, clip=true, width=0.25\columnwidth]{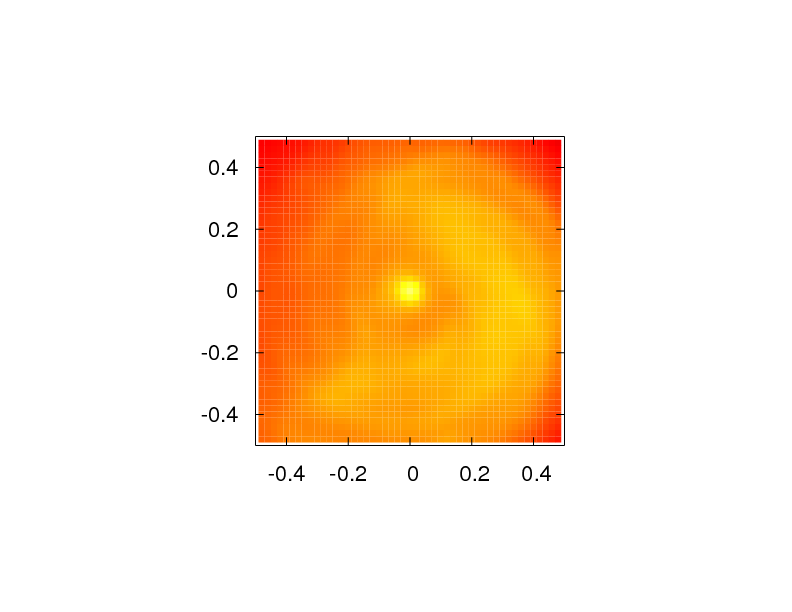}}
        &
       \raisebox{-2cm}{  \includegraphics[trim=7cm 4.3cm 7cm 4.3cm, clip=true, width=0.25\columnwidth]{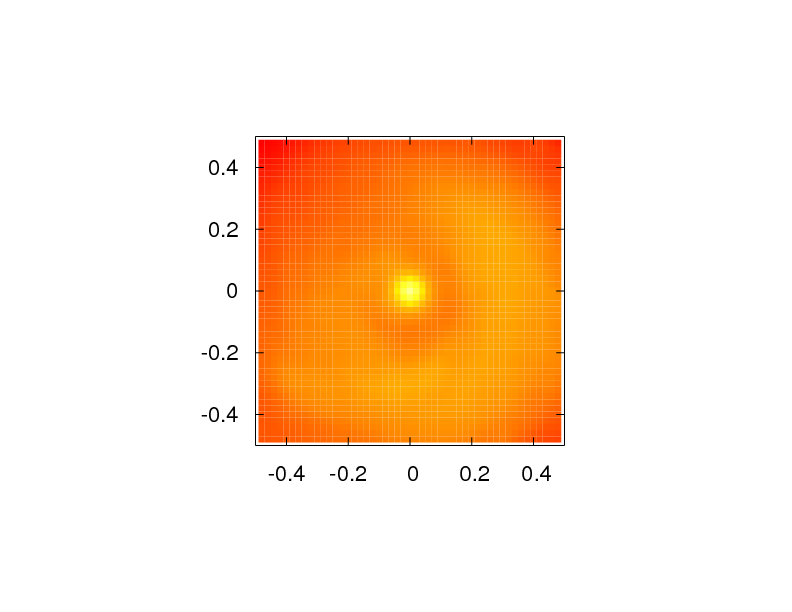}}
        \\  & $\hat{x}$ & $\hat{x}$ & \hspace{2cm}$\hat{x}$ 
\end{tabular}
        \caption{Projected surface densities (in log scale) for No-kick, and kick angles of 15, 30, 45 and 60 degrees (rows from top to bottom respectively); at $\hat{t}_* = 4, 8, 12$.     \label{fig:surfDens-map_zoomIN}}
\end{center}
     
\end{figure}

While the azimuthal averages provide a clear picture of the global behavior of the disk, they do mask some of the more local phenomenon that develop after the kick. In Figs.~\ref{fig:surfDens-map} and \ref{fig:surfDens-map_zoomIN} we show colormaps of the surface density at larger and smaller scales, respectively, with different color mappings between the two to allow for maximum contrast.   The large scale maps show the global expansion of the disk,  though some care is required to interpret the results for the more oblique kicks. Most of the material that remains bound after the kick falls inward quickly to form the circularized inner disk extending out to $\hat{r}\lesssim 2.0$.  The lower arc that appears clearly for the $45^\circ$ and $60^\circ$ on the left (smaller $x'$-values)  of the inner disk (and more subtly so for the $30^\circ$ case) represents the unbound upper portion of the original disk (i.e., the material initially near $\phi=90^\circ$), with the density enhancements primarily a 2-dimensional projection effect.  The material lies below (i.e., at lower $z'$-values) that the bound component of the disk, except for the marginally unbound material where the stream connects to the inner disk.  The upper arc visible clearly in the more oblique cases represents the marginally bound component of the disk from the upper edge of the original disk.  This stream of material initially moves way from the BH after the kick while remaining level vertically with the inner disk before accreting toward the inner disk following a roughly ballistic trajectory.  When it collides with the inner disk, it shock heats and circularizes, with an accretion rate that decreases gradually over time. 
In each of our runs, the shock fronts are never particularly sharp, certainly less so than the adiabatic 2-d thin-disk calculations in \citet{Corrales:2009nv} which are themselves much more spread out than isothermal calculations in which there is no shock heating.  Instead, because we allow the gas to heat as it shocks, the spiral patterns rapidly blur, leading to more extended density enhancements.  The non-axisymmetry is strongest during the early phases of the simulation, and gradually fades as the disks relax and collisions circularize the fluid, so that by $\hat{t}_*=12$ we see only minor deviations from axisymmetry, particularly near the outer edge of the bound region where infalling matter is still playing a role.

In the smaller-scale surface density plots, Fig.~\ref{fig:surfDens-map_zoomIN}, the role of the ``gap'' at $\hat{t}_*=4$ is immediately apparent.  For the $60^\circ$ kick, the center of the disk has already filled in, though the surface density is strongly non-axisymmetric even at very small radii.  Significantly more matter is located at small separations for the $45^\circ$ kick simulation, but a hollow is clearly visible around the BH.  For the more vertical kicks, the gap is clearly present and very little matter is evident to begin filling it.  By $\hat{t}_*=8$, this influx of matter leads to a very sharp increase in the central density for the more oblique kicks that is not present in the more vertical ones.  Finally, by $\hat{t}_*=12$, the disk exhibits a much greater degree of axisymmetry, with only the most oblique kick case, in which the bound component of the disk is drawn almost entirely from one side of the pre-kick accretion disk, still retaining a marked angular dependence pattern.

\begin{figure}
\begin{center}
	\includegraphics[trim=0cm 10cm 0cm 0cm, clip=true, width=\columnwidth,height=0.35\columnwidth]{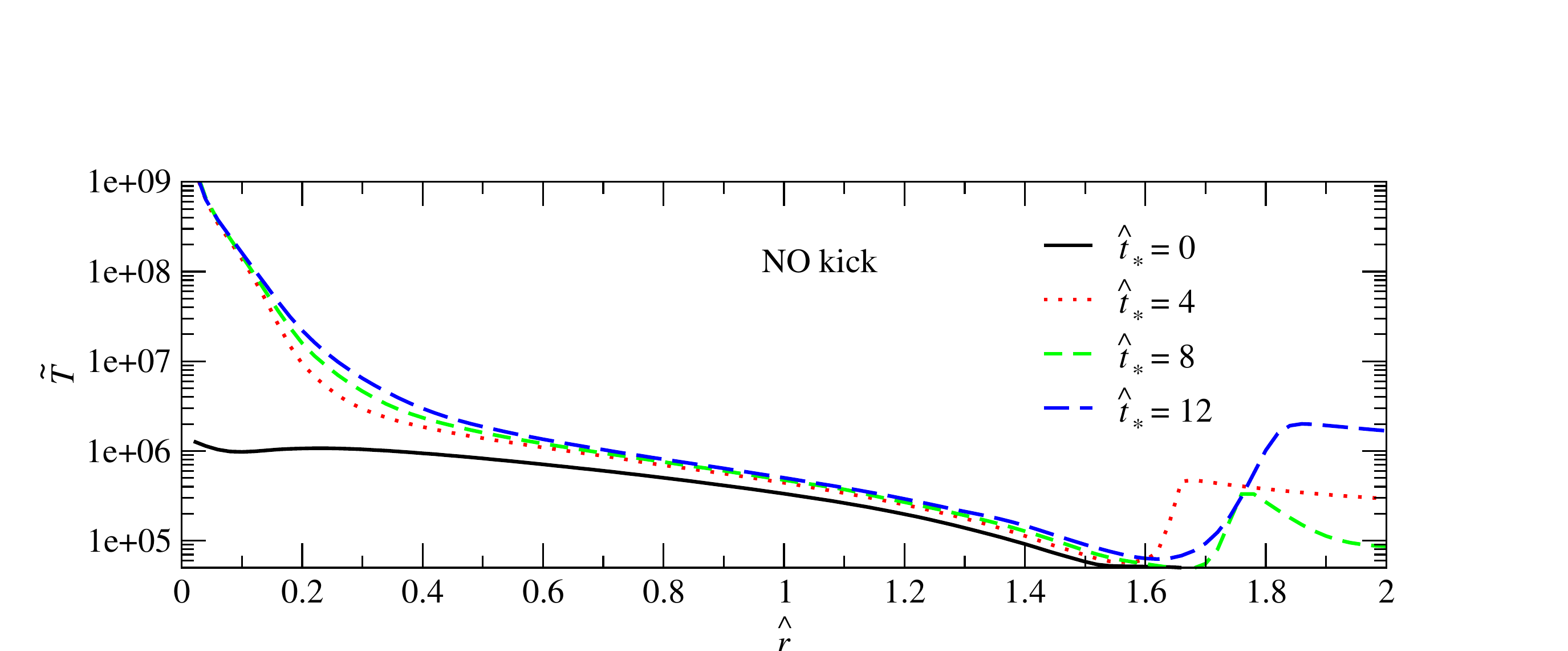}
        \vspace{-2cm}
	\includegraphics[trim=0cm -1cm 0cm 2cm, clip=true, width=\columnwidth]{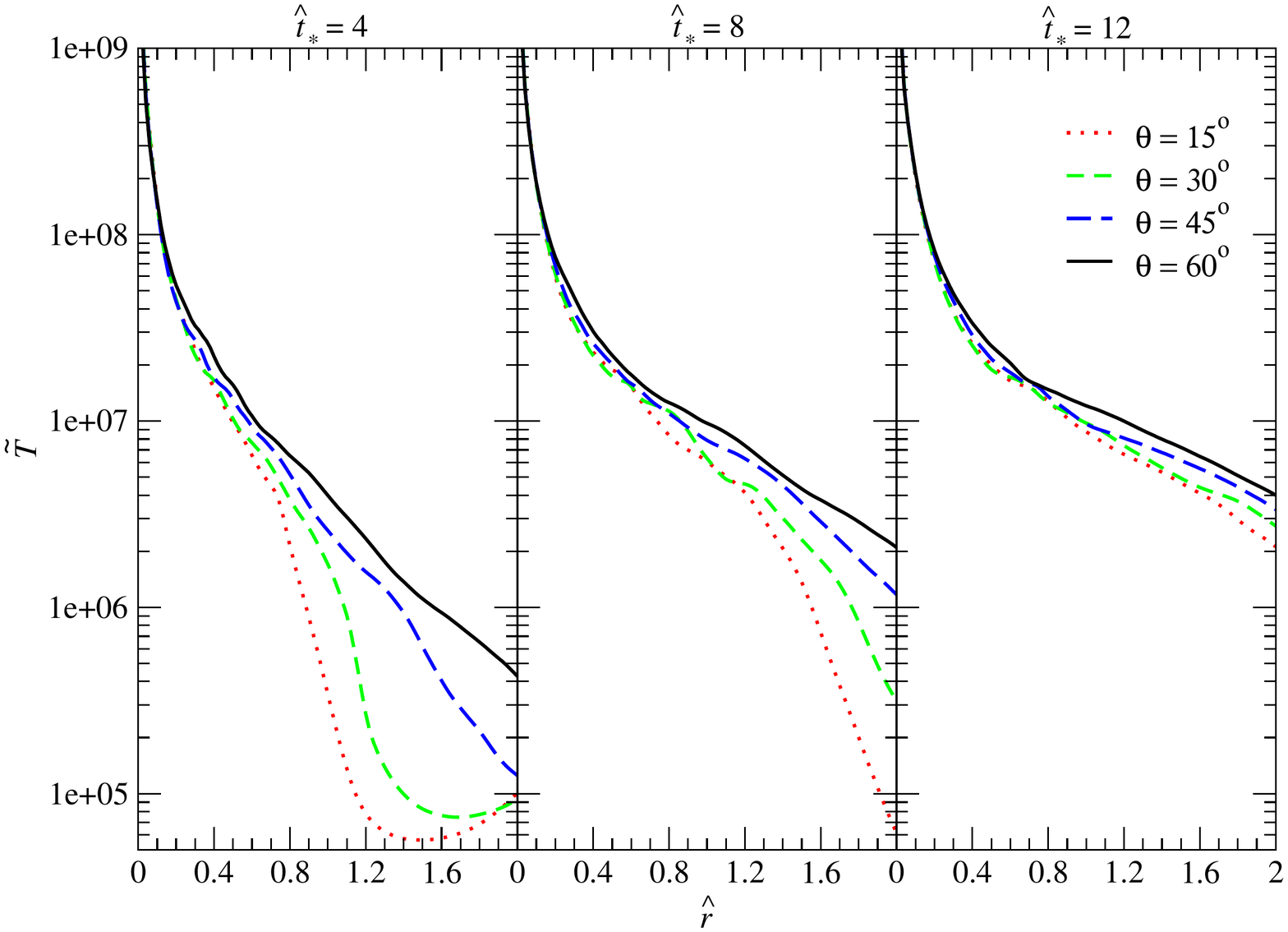}
        \caption{Radial profiles (averaged over azimuthal angle) of temperatures (in log scale) for  the no-kick model (top panel), and kick angles of 15, 30, 45 and 60 degrees; at $\hat{t}_*=0, 4, 8, 12$ (from left to right).   \label{fig:Temp-rad_profiles}
}
\end{center}
      \end{figure}

Turning to the thermal evolution of the disks, we see in the first panel of Fig.~\ref{fig:Temp-rad_profiles} the radially averaged thermal profile of our adiabatic initial model, for which $T\propto \rho^{2/3}$.   We find substantial heating very quickly in the center of the disk, with strong shocking throughout the region $\hat{r}\lesssim 1$.  The strong shock heating extends out through the majority of the bound component of the disk, except for the outermost regions at early times.  Indeed, the sudden dropoffs in the temperature profiles corresponding to the more vertical kicks ($15^\circ$ and $30^\circ$) at $\hat{t}_*=4$ occur toward the outer part of the bound region of the disk, not the unbound part of the disk.  This is in accordance with our previous discussion, as the further reaches of the bound components for the more vertical kicks are initially sent outward in their orbits, and take longer to eventually collide with other fluid streams and shock.  Over time, the temperature profile smooths out, and by $\hat{t}_*=12$ we have essentially a single temperature profile, peaked in the center and then falling off more slowly at larger radii, that characterizes all of our kicked disk simulations.

Based on the surface density and temperature profiles of our simulations (Figs.~\ref{fig:surfDens-rad_profiles} and \ref{fig:Temp-rad_profiles}), it becomes clear that the differences in the global energies of the disks as a function of the kick angle are not the result of radically different temperature profiles, nor significantly different densities throughout the bulk of the disk, but rather from density differences in the innermost region of the disk.  It is at small radii that all three energies take on their largest magnitudes, and the factor of $5-10$ difference in the surface densities at these radii represent a significant fraction of the total energy, though not the total mass.  Perhaps the clearest prediction from our simulations is that we expect oblique kicks to produce much more energetic signatures and substantially higher measured accretion rates for timescales of thousands of years for systems similar to our reference model and almost an order of magnitude longer for systems with kicks of roughly $500$km/s.

\begin{figure}
\begin{center}
	\includegraphics[width=0.85\columnwidth]{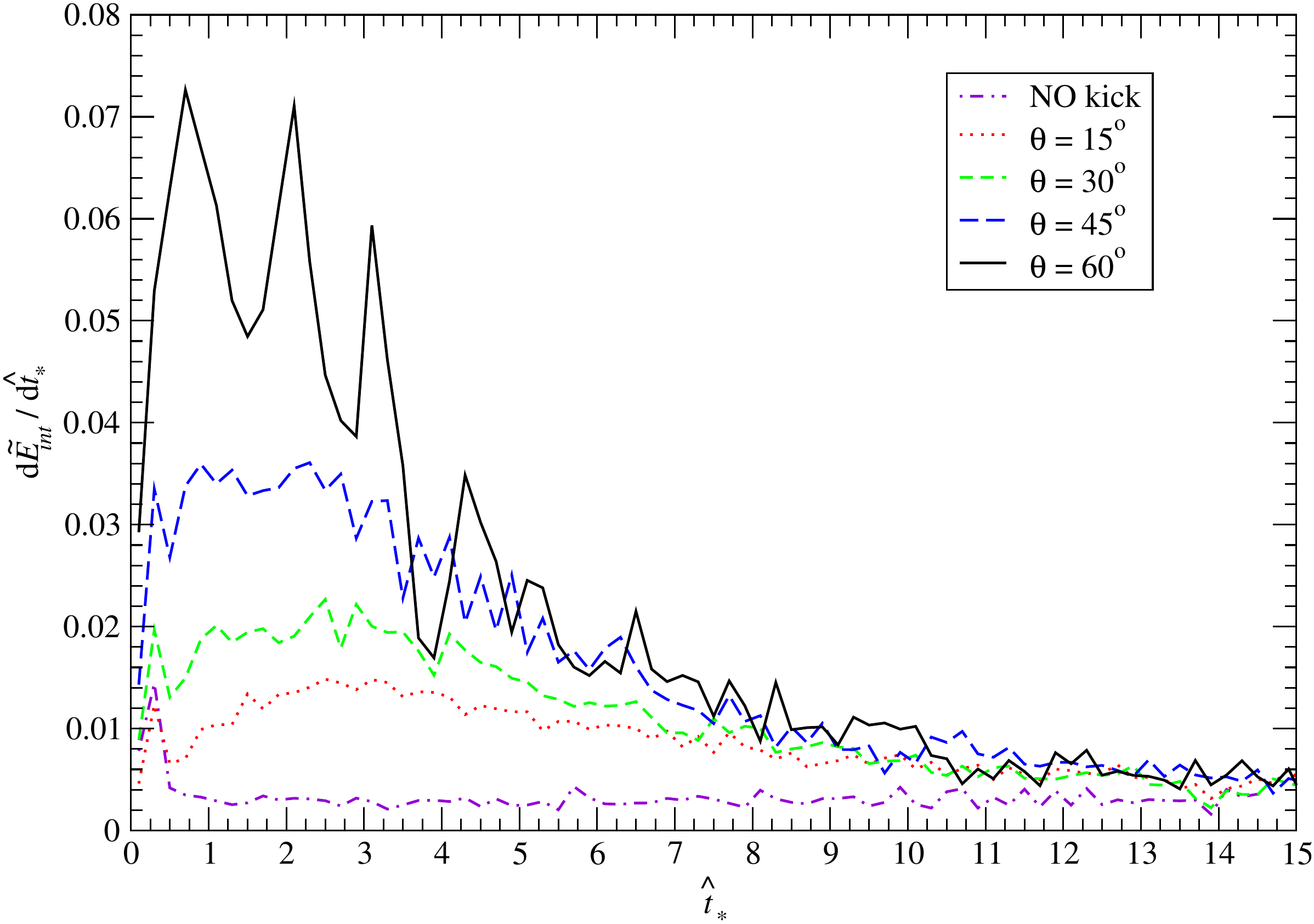}
	\caption{Potential disk luminosities from our disk models, calculated as the time derivative of the internal energy as shown in Fig.~\protect\ref{fig:energies}.  To smooth the data and minimize SPH discretization noise, we difference over intervals of 500 (No kick, $15^\circ$ and $30^\circ$) or 1000 timesteps ($45^\circ$ and $60^\circ$).  Besides the expected increase on average luminosity as the kick becomes more oblique, we also see substantial oscillations in luminosity for the more oblique kicks, especially the $60^\circ$ case.}        \label{fig:Luminosity}
\end{center}
\end{figure}

Our dynamical treatment assumes the gas is optically thick, so that the disk is allowed to shock heat without radiative cooling applied.  Nevertheless, if we estimate the potential luminosity of the disk by assuming that the internal energy gains would be immediately radiated away but not affect the dynamics in any other way, we can convert the internal energy profiles from Fig.~\ref{fig:energies} into a luminosity by assuming that $\left(\frac{dE}{dt}\right)_{rad} = \frac{dE_{INT}}{dt}$.  The results are shown in Fig.~\ref{fig:Luminosity}, where we differenced over intervals $\delta t_*=0.2$ to minimize spurious noise. In all cases, we see an immediate but purely numerical luminosity peak when we end relaxation and begin the dynamical evolution.   For the unkicked disk, we see only a very small, nearly constant luminosity over time.   For the more vertical kicks, we find a small rise in luminosity followed by a gradual decline, but with very little temporal structure.  For the more oblique kicks, we see both significantly larger luminosities as well as quasiperiodic emission spikes, especially for the $60^\circ$ case, in which there are persistent oscillation amplitudes of tens of percent with a period of slightly longer than $\hat{t}=1$.  Given the density patterns we observed, it seems clear that we are observing periodic shocking due to intersecting flows followed by accretion events as dense regions within the inner disk fall toward the BH while heating up significantly.  The timescale roughly corresponds to the orbital timescale at the inner edge of the disk, with the strong $m=1$ mode dependence of the fluid density (i.e., a single-arm spiral pattern) leading to increased shocking when the pattern wraps a full time around the BH.

\begin{figure}
\begin{center}
	\includegraphics[width=0.85\columnwidth]{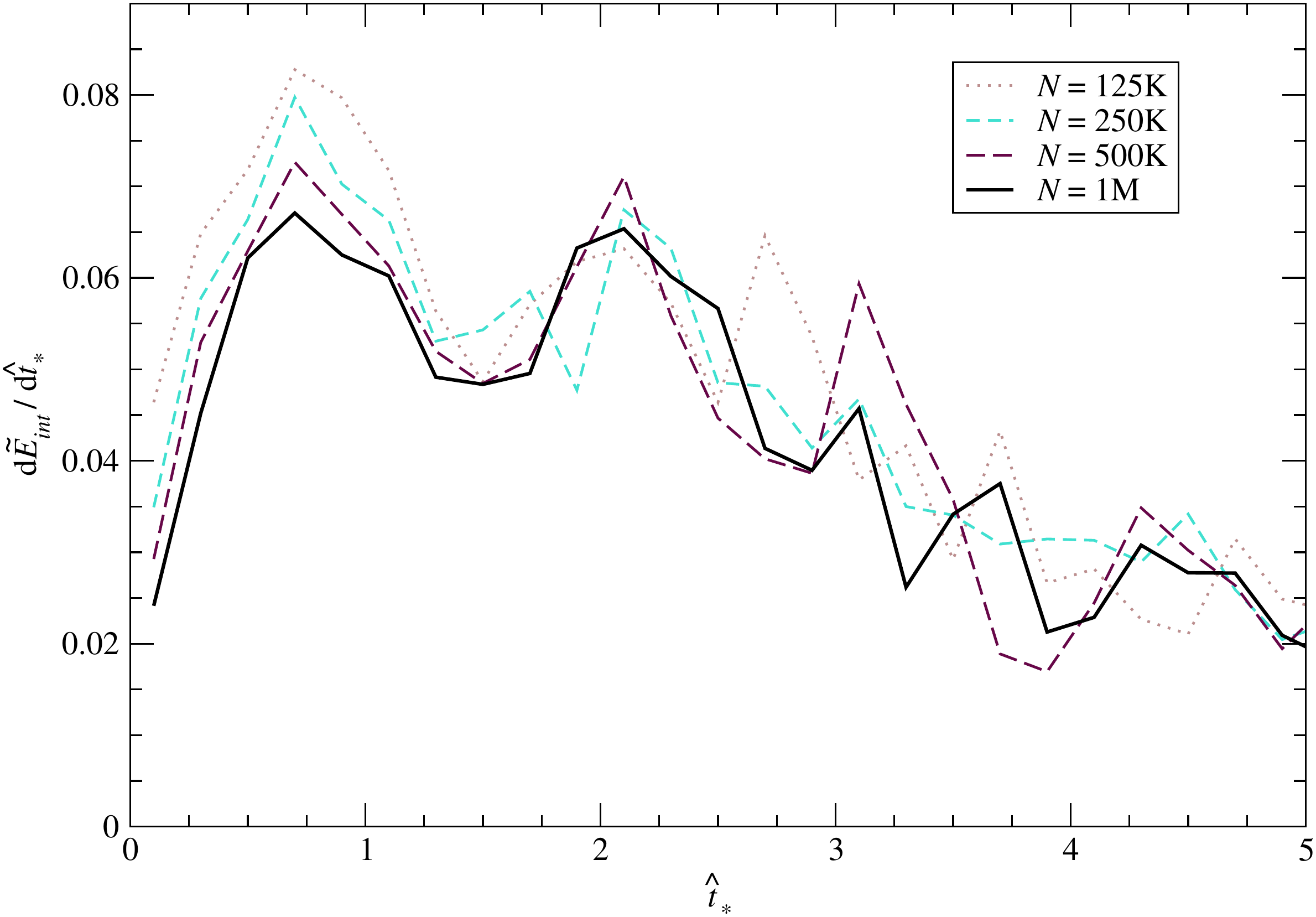}
	\caption{Disk luminosities from our $60^\circ$ cases with varying numbers of SPH particles, showing the results for $N=10^6$ (solid), $5\times 10^5$ (long-dashed; this is the same run shown in Fig.~\protect\ref{fig:Luminosity}), $2.5\times 10^5$ (short-dashed) and $1.25\times 10^5$ (dotted) SPH particles.  We see clear evidence for two peaks in the luminosity profle, at roughly $\hat{t}_*=0.7$ and $2.1$, and possible evidence for further modulation of the signal at later times.}
	      \label{fig:Lconverge}
\end{center}
\end{figure}

To confirm that these results are robust, and not merely a numerical artifact, we re-ran the $60^\circ$ case using $1.25\times 10^5$, $2.5\times 10^5$, and $10^6$ SPH particles in order to compare to our $N=5\times 10^5$ reference model.  The luminosity we derive from each run is shown in Fig.~\ref{fig:Lconverge}.  We see good agreement throughout the early phases, with slightly lower luminosities during the first peak for higher particle numbers because spurious numerical dissipation is slightly smaller.  The drop in luminosity at $\hat{t}_*=1$ is a robust effect, as is the second luminosity peak at $\hat{t}_*=2$.  There is evidence for further oscillations in the luminosity, but not at a level we feel is conclusive.  Still, given the strong convergence of the numerical results, it seems clear that the quasi-periodic luminosity is a physical prediction of our model, and not a numerical artifact owing to under-resolution of the accretion flows.

\subsection{Prompt radiation calculations and comparison to previous work}\label{sec:noheat}

Our assumption of Sec.~\ref{subsec:heat} that the disk is allowed to shock heat differs from most previous works, particularly \cite{Corrales:2009nv} and \cite{Rossi:2009nk}.  We note that the assumption is closer to reality than it is often given credit for; indeed, while our default parameters from \S\ref{subsec:scales} indicates that the initial disk is optically thin, faster kicks, larger disk masses, or smaller black holes can potentially tip the balance toward optical thickness across much of the disk: see Eq.\ \ref{eq:depth} and Fig.~\ref{fig:surfDens-rad_profiles}.  In the innermost regions of the disk, where the majority of the luminosity actually arises, optically thick regions do develop, and it seems clear that immediate and passive cooling itself neglects important physical contributions to the dynamics of the disk.
In order to bookend the likely observable emission from disks, we have also run simulations using a ``prompt radiation'' formalism, in 
which the energy evolution equation, Eq.~\ref{udot}, is ignored, and we instead assume that all energy is immediately radiated according to Eq.~\ref{eq:eintimplied}, resulting in an adiabatic evolution.  

Calculations we perform using the prompt radiation assumption indicate that if the material does not get shock-heated, there are severe numerical challenges in evolving the inner region even when the BH potential is softened.  Indeed, given the lack of thermal pressure support, we find generically that particles accumulate catastrophically around the BH, leading to a rapid slowdown of the evolution when the Courant timestep grows too small.  Prior SPH works in which a prompt radiation assumption was used (see, e.g., \cite{Rossi:2009nk}) have implemeted an accretion radius $R_{acc}$, inside of which all particles are immediately accreted and removed from the simulation.  In \cite{Rossi:2009nk}, the accretion radius was set to the inner edge radius of the initial disk, $R_{acc}=0.1$ (G.\ Lodato, private communication).

\begin{figure}
\begin{center}
	\includegraphics[width=0.85\columnwidth]{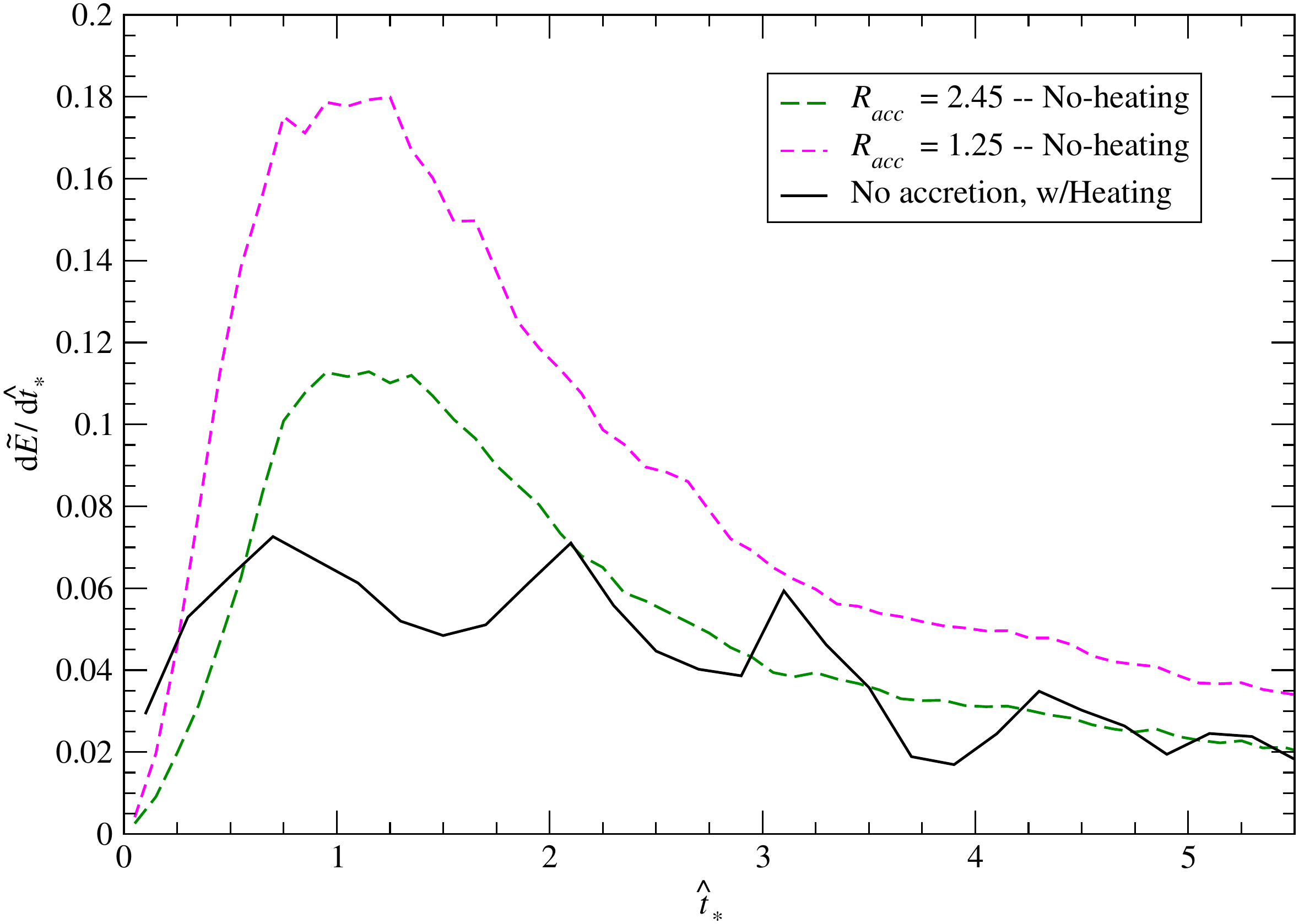}
	\caption{Disk luminosities from our $60^\circ$ case with the effects of shock heating included (solid curve), compared to two ``prompt radiation'' runs in which shock heating is ignored and luminosities calculated implicitly from Eq.~\protect\ref{eq:eintimplied}.  For the latter, we impose an accretion radius $R_{acc}$ inside which all particles are assumed to be accreted by the BH and removed from the numerical simulation, choosing $R_{acc}=0.098$ (long-dashed, just inside the inner edge of the original pre-kick disk) and $R_{acc}=0.05$ (short dashed, at half that separation).  We see substantially larger luminosities for the prompt radiation cases, but with a strong dependence on the choice of the accretion radius, which is not a physically motivated parameter but rather a numerically chosen one.}  
	      \label{fig:Lnoheat}
\end{center}
\end{figure}

To allow a more direct comparison with \cite{Rossi:2009nk},
we have run two prompt radiation models, using a $60^\circ$ kick , setting the accretion radius to be $R_{acc}=0.098$, just inside the inner edge of our initial disk, and $R_{acc}=0.05$, at half that radius. (We note that our angular convention is reversed from theirs: we refer to a vertical kick as $\theta=0^\circ$ and an in-plane kick as $90^\circ$.)
For such runs, strict energy conservation is impossible to maintain since the density contribution from accreted particles toward other particles is instantaneously removed in the middle of each timestep.  Still, it is straightforward to calculate the implied luminosity rate, which we show in Fig.~\ref{fig:Lnoheat} for the two prompt radiation runs and our original run from Sec. \ref{subsec:heat} with shock heating included.

Adding an accretion radius has several rather dramatic effects on the resulting luminosity.  It smooths out the predicted luminosity, since high-energy interactions are suppressed, and increases the minimum timestep since particles never approach as closely to the BH.  It also introduces an unphysical scale into the results, which becomes clearly evident when we compare results from the $R_{acc}=0.098$ and $R_{acc}=0.05$ cases.  Indeed, we see an increase of over 50\% in the predicted luminosity when the accretion radius is halved, since the characteristic energies of particles immediately before accretion, which represent a substantial component of the energy release, are roughly a factor of two greater based on dimensional arguments.  As a result, conclusions about the potential luminosity of the disk unfortunately depend strongly upon the ad hoc choice of accretion radius.  Also, we note this produces an unphysical accretion-driving process, since pressure support from the inner disk is removed resulting in strongly anisotropic pressure support for particles just outside the accretion radius.  This leads to a persistent dependence in the long-term luminosity on the accretion radius, which is worrisome. 

A secondary effect of implementing an accretion radius is to eliminate periodicity in the luminosity of the disk for kicks of high inclination angles, which we believe results from excising from the computational domain the inner region of the disk where accretion flows with the highest energies typically interact.  This is a problem for the 60 degree case in particular, as the post-kick periastron distances fall within the accretion radius for much of the initial bound component of the disk.  

We can now classify the effects of both shock-heating and inclusion of the innermost disk, as in our simulations of section \ref{subsec:heat}, on the results of numerical simulations.
 Here, we see much more modulated emission produced from oblique kicks, as a result of high-eccentricity flows into the inner disk, while more vertical kicks yield much smoother luminosity profiles.  In contrast, \cite{Rossi:2009nk} 
  found substantial oscillations for a vertical kick, which we suggest might result from modulations in the accretion rate into the inner disk but exclusion of the emission that would otherwise take place within $\hat{r}\lesssim 0.1$.
  Our luminosities also peak at later times than their simulations indicated (see their Fig.~20).  Among the possible explanations, the likeliest is that our simulations are most sensitive to colliding flows in the inner regions of the simulation, which require infalling material, whereas the simulations of \cite{Rossi:2009nk} will have accreted much of the inner disk by the time such flows reach the center.  As a secondary issue, 
 the different initial disk configurations may also play a role, as we assumed an equilibrium model while theirs followed a power-law profile.  The issue of adiabatic versus isothermal simulations also deserves mention, though spiral shocks are often stronger in isothermal models \citep{Corrales:2009nv}.  
 
 There is a precedent for the periodic luminosities we see, as a very similar profile appears in Fig.~4 of \cite{Shields:2008va}, who studied a collisionless model for kicked disks that modeled emission via passive ``collisions'' assumed to occur when particles approached close to each other, though without any feedback on the resulting dynamics.  They also found two strong luminosity peaks and a quasi-periodic luminosity profile afterwards, though for a $45^\circ$ kick (we note that their disk model was different than ours, as it lacked pressure support, and that the normalization of time between their runs and ours differs by a factor of $2\pi$).  This strongly suggests that the kick luminosity in our simulations is driven by interacting flows, even in the presence of disk heating, and that these flows are not being modeled in simulations that excise the inner region of the disk.
 
Although our initial disk model differs from that of \citet{Rossi:2009nk}, it still provides an opportunity to check our results alongside the ``circularization'' model proposed there, in which one calculates the approximate available energy for a fluid element within the disk by using its post-kick specific angular momentum $l$ to infer a circularization radius $r_c=L^2/GM$ and energy $-GM/(2r_c)$, and subtracting this away from the fluid element's initial energy after the kick (see their Eqs.~6 -- 10).    Rather than construct the energy estimate by a surface integral over the disk, we compute it particle-by-particle from the common disk configuration present immediately after relaxation.   Our results are shown in Table~\ref{table:energies}, where the columns list the predicted ``circularization energy'' $\tilde{E}_c$, the actual internal energy $\tilde{E}_{\rm INT}$ at  $\hat{t}_*=12$, and the difference between the internal energy of a kicked model and our unkicked reference model $\tilde{E}_{{\rm INT};kick}\equiv \tilde{E}_{\rm INT}-\tilde{E}_{{\rm INT};nokick}$ at $\hat{t}_*=12$, for each of the kicked runs.
We see that while the overall values are relatively close, indicating that the circularization model yields a good approximation for the order of magnitude of the energy that could be emitted by the disk, this very simple model does not yield quantitatively accurate predictions with respect to the kick angle dependence.
Thus, while the formula is certainly useful for establishing the approximate disk luminosity given a reasonable timescale for emission, we do not see strong evidence that it can be extrapolated to angles that lie very close to the disk plane, for which the predicted energy 
rises  like a power law in the out-of-plane angle ($90^\circ-\theta_k$ in our notation)  to extremely large values (see Figs.~4 and 21 of \citet{Rossi:2009nk}).

\begin{table}[ht!]
\begin{center}
\caption{Estimated energies available to the disk \label{table:energies}}
\begin{tabular}{cccc}\hline\hline
Kick angle $(^\circ)$ & Circularization energy & Internal energy at $\hat{t}_*=12$ &  Internal energy, corrected for kick\\ \hline
15 & 0.13 & 0.12 &0.07 \\
30 & 0.11 & 0.16 & 0.11 \\
45 & 0.19 & 0.24 & 0.19\\
60 & 0.48 & 0.32 & 0.27 \\ \hline
\end{tabular}
\end{center}

\end{table}

\section{Discussion and future work}\label{sec:discussion}

In this paper, we have studied the response of a quasi-equilibrium accretion disk around a SMBH that undergoes an impulsive kick, presumably because of a binary merger and the corresponding asymmetric emission of linear momentum in gravitational waves.  While there are a number of sources that have been identified as candidate kicked disks with recoil velocities so large that they would be in the far tails of the kick velocity distribution, our results here are scale-free with regard to the kick, and may be applied to a broad swath of potential kicked systems.  Indeed, while for our assumed reference model with $M_{\rm BH}=10^8M_\odot$, $m_{disk}=10^4M_\odot$, and $v_{\rm kick}=1000$ km/s we find characteristic timescales of roughly 1000 years and disk luminosities and temperatures of up to $10^{42}$erg/s and $10^{8-9}$K, respectively, the results should work for a wide range of masses and kick velocities.
There is a trade-off, to be had, of course, especially given the dependence of many of the quantities we investigate on the kick velocity.  Indeed, cutting the kick velocity by half and leaving the masses of the BH and disk unchanged increases the timescales we consider by a factor of eight while cutting the energies and temperatures by a factor of four and thus the luminosities by 32, indicating that there should be a definite observational bias toward the larger kicks.

The code we introduce uses 3-dimensional SPH techniques, and is modified to incorporate a Lagrangian (``grad-h'') prescription that allows energy to be conserved to high precision over the course of all our runs.  We also introduce a Lagrangian-based black hole smoothing potential, which proves critical in allowing us to avoid numerical issues associated with point-like potentials.   One of the most important purely numerical conclusions of this work is that black holes must be handled extremely carefully in SPH, since they can introduce particle clustering instabilities that are both conservative but highly unphysical.  This treatment, along with the inclusion of shock-heating, has allowed us to perform what we believe to be the first 3-d SPH simulations of these systems that do not require imposing an ad hoc ``accretion radius'' inside of which SPH particles are removed from the simulation.

In order to examine the phase space of post-kick disks, we have varied the kick angle of the SMBH with respect to the initial disk orientation, which has a large effect on the resulting evolution.  More oblique kicks, i.e., those most oriented toward the equatorial plane of the original disk, produce substantially higher peak luminosities for a given BH and disk mass and kick velocity, with roughly a factor of four gain between kicks oriented $15^\circ$ away from vertical and those oriented at $60^\circ$ from vertical.  Assuming an astrophysical context in which the disk is aligned with the SMBH binary orbit prior to merger, it is unlikely that more oblique kicks will actually yield more luminous events, however.  Kick velocities are systematically higher for more vertical kicks \citep{Lousto:2009ka}, and given the $v_{kick}^5$ dependence of the luminosity, the brightest kicked disks are likely to be those with the highest speed kicks, with the kick angle playing a secondary role.

We find more rapid luminosity peaks appearing for more oblique kicks, which is the opposite result from a previous set of simulations that considered power-law density profile disks \citep{Rossi:2009nk}. Based on our $60^\circ$ calculation, we attribute this to the rapid filling of the innermost region of the disk by material kicked into high eccentricity orbits with extremely small periastron distances (see Fig.~\ref{fig:rperi}), which flows inward from the inner edge of the pre-merger disk; we note that this pattern differs from that found in Fig.~15 of \citet{Rossi:2009nk}, likely due to the combined effects of shock heating and their excision of particles from the innermost regions of the disk.   Clearly, one of the important conclusions of this work and others is that flows of material toward the SMBH after the kick can release tremendous amounts of energy, but need to be modeled very carefully to derive reasonable light curves and spectra, pushing the limits of current numerical simulations.

The surface density oscillations and the resulting quasi-periodic emission we observe in time for the most oblique kicks will make an interesting
topic for further research.  Based on our ``gap-filling'' model  it seems clear the dynamics of the inner disk play an important role in the dynamical evolution and EM emission from the disk, and it would be highly useful to extend the same simulations on both sides of the kick to explore the full history of a circumbinary/post-kick disk. In particular, it will be important to accurately resolve the inner edge of the circumbinary disk, where the tidal field of the binary should have swept out a gap \citep{Schnittman:2008ez}, and to understand how infalling material from that region interacts with the more tenuous tidal streams of matter in the process of accreting onto the SMBH prior to their merger \citep{Macfadyen:2006jx,Hayasaki:2006fq,Kimitake:2007fs}. It will also be useful to incorporate a more thorough treatment of the radiative evolution of the disk, since a proper treatment of radiative cooling and disk opacities will break the scale-invariance of the results and allow for much more accurate predictions of observable phenomena.

\acknowledgments

We thank Jeremy Schnittman for useful conversations including pointing out the importance of the kick angle-kick velocity relationship, and Fabio Antonini, Manuela Campanelli, Evgenii Gaburov, Julian Krolik, Carlos Lousto, David Merritt, Scott Noble, and Yosef Zlochower for helpful discussions.  J.A.F. acknowledges support from NASA under  awards 08-ATFP-0093 and HST-AR-11763.01.  M.P. acknowledges support from the NSF under award PHY-0929114.   This research was supported in part by the National Science Foundation through TeraGrid resources provided by NCSA and TACC under grant numbers TG-PHY060027N and TG-AST100048
and has made use of both the SPLASH visualization software \citep{Price:2007he} and NASA's Astrophysics
Data System.

\begin{appendix}

\section{SPH Evolution scheme}

Our SPH evolution scheme uses a variable smoothing length approach following the formalism described in \citet{Springel:2001qb} and \citet{Monaghan:2001MNRAS.328..381M}.
Defining a particle-based parameter
\begin{eqnarray*}
\Omega_i = 1-\frac{\partial h_i}{\partial \rho_i}\sum_j m_j\frac{\partial W_{ij}(h_i)}{\partial h_i},
\end{eqnarray*}
our acceleration equation is $\dot{\bf v}_i=\dot{\bf v}_i^{({\rm SPH})}+
\dot{\bf v}_i^{({\rm grav})}$ with
\begin{equation}
\dot{\bf v}_i^{({\rm SPH})} = -\sum_j m_j \left[
\left(
      \frac{P_i}{\Omega_i\rho_i^2}+\frac{\Pi_{ij}}{2}
\right) {\bf \nabla}_i W_{ij}(h_i)+
\left(
      \frac{P_j}{\Omega_j\rho_j^2}+\frac{\Pi_{ji}}{2}
\right) {\bf \nabla}_i W_{ij}(h_j)
\right]\,,
\label{fsph}
\end{equation}
while our evolution equation for the entropic variable $A_i$ is
\begin{equation}
{d\,A_i\over d\,t}={\gamma-1 \over \rho_i^{\gamma-1}}\sum_jm_j
      \frac{\Pi_{ij}}{2}
({\bf v}_i-{\bf v}_j)\cdot{\bf\nabla}_iW_{ij}(h_i)\,.
\label{udot}
\end{equation}
For the ``prompt radiation'' simulations discussed in Sec.~\ref{sec:noheat}, we hold $A_i$ fixed and ignore Eq.~\ref{udot}, though we do evolve Eq.~\ref{fsph} as written.  Energy losses are calculated by integrating the implied change in the entropic variable over time $\frac{dA_i}{dt}$, and inserting this into the expression
\begin{equation}
\frac{dE_{implied}}{dt}  = \sum_i \frac{1}{\gamma-1} m_i (dA_i/dt)_{implied} \rho_i^{\gamma-1} =
\sum_i \sum_jm_im_j
      \frac{\Pi_{ij}}{2}
({\bf v}_i-{\bf v}_j)\cdot{\bf\nabla}_iW_{ij}(h_i)\,.
\label{eq:eintimplied}
\end{equation}

In the above, $W_{ij}(h_i)$ is the normalized second-order accurate kernel function introduced in \citet{MonaghanLattanzio85}, used widely throughout the SPH community, and $h_i$, $\rho_i$, $m_i$, and $P_i$ are the SPH particle smoothing lengths, densities, masses, and pressures, respectively.  The artificial viscosity term $\Pi_{ij}$ is discussed below.  We assume a smoothing length-density relation in the form
\begin{equation}
h_i = \left(\frac{1}{h_{max}}+b_i\rho_i^{1/3}\right)^{-1} \label{hofrho}
\end{equation}
where $h_{max}=50$.  The $b_i$ values are chosen so that each particle should have $\sim 200$ neighbors given the initial density profile of the disk and are updated to maintain this condition during relaxation.  Once the dynamical phase of the evolution begins, we hold $b_i$ fixed and  solve implicitly at each time step for the proper smoothing length and density that satisfy equation (\ref{hofrho}).

Because the self-gravity of the disk is ignored here, the only
gravitational force acting on the particles comes from the black hole.
We assume instantaneous Newtonian gravitation, neglecting retardation
effects from the moving black hole since our characteristic speeds are
small fractions of the speed of light.   Although we treat the black
hole as a pure point mass without any {\it intrinsic} softening, the
gravitational force on any SPH particle within two smoothing lengths
of the black hole is softened according to the mass distribution of that SPH
particle itself, as described by its smoothing kernel.
To do so, we
follow the formalism of \citet{Price:2006iz}, who use a variational approach
to derive equations of motion that properly account for variable smoothing lengths.
In particular, we start by writing the gravitational part of the Lagrangian as
\begin{equation}
L_{\rm grav}=-\sum_i m_i \Phi_i = -G M_{\rm BH} \sum_i m_i \varphi_i(h_i). \label{Lgrav}
\end{equation}
The last equal sign in equation (\ref{Lgrav}) defines the gravitational potential $\Phi_i$ of particle $i$.
Here $\varphi_i(h_i)$ refers to $\varphi(r_i,h_i)$ where $r_i=|{\bf r}_i|$ is the distance of particle $i$ from the BH and
\begin{equation}
\varphi(r,h) = \left\{ \begin{array}{ll}
\left(- \frac{7}{5} + \frac{2}{3}q^2 - \frac{3}{10}q^4 + \frac{1}{10}q^5 
 \right)/h, & 0 \le q < 1; \\
\left(- \frac{8}{5} + \frac{1}{15q} + \frac{4}{3}q^2 - q^3 + \frac{3}{10}q^4 - \frac{1}{30}q^5 
 \right)/h, & 1 \le q < 2; \\
-1/r & q \ge 2 \end{array} \right., \label{eq:cubicsplinepotential}
\end{equation}
with $q=r/h$, is the gravitational potential associated with the usual SPH smoothing kernel \citep[e.g.][]{Price:2006iz}.

Following the approach of \S3 of \citet{Price:2006iz}, but with our Lagrangian,
we find
\begin{equation}
\dot{\bf v}_i^{({\rm grav})} =-G M_{\rm BH}\nabla_i\varphi_i(h_i)
-G M_{\rm BH}\sum_j m_j \left[
      \frac{1}{\Omega_i}\frac{\partial \varphi_i}{\partial h_i}\frac{\partial h_i}{\partial \rho_i}
{\bf \nabla}_i W_{ij}(h_i)+
      \frac{1}{\Omega_j}\frac{\partial \varphi_j}{\partial h_j}\frac{\partial h_j}{\partial \rho_j}
{\bf \nabla}_i W_{ij}(h_j)
\right]. \label{vdotgrav}
\end{equation}
The first term on the right hand side of equation (\ref{vdotgrav}) is the usual softened gravitational acceleration, while the remaining terms allow for variable smoothing lengths and preserve energy conservation.  We note that one of these correction terms vanishes when the kernel of the relevant SPH particle does not overlap with the BH, because, given equation (\ref{eq:cubicsplinepotential}), $\partial \varphi/\partial h=0$ whenever $r>2h$.
Summing equations (\ref{fsph}) and (\ref{vdotgrav}), the total acceleration of particle $i$ is calculated as
\begin{equation}
\dot{\bf v}_i = -G M_{\rm BH}\nabla_i\varphi_i(h_i) -\sum_j m_j \left[
\Upsilon_{ij} {\bf \nabla}_i W_{ij}(h_i)+
\Upsilon_{ji} {\bf \nabla}_i W_{ij}(h_j)
\right]\,, \label{vdottotal}
\end{equation}
where
\begin{equation}
\Upsilon_{ij}=
      \frac{P_i}{\Omega_i\rho_i^2}+\frac{\Pi_{ij}}{2}+\frac{G M_{\rm BH}}{\Omega_i}\frac{\partial \varphi_i}{\partial h_i}\frac{\partial h_i}{\partial \rho_i}
\end{equation}
and where, given equation (\ref{hofrho}), we use $\partial h_i/\partial \rho_i=-b_ih_i^2/(3\rho_i^{2/3})$.

We implement the AV form
\begin{equation}
\Pi_{ij}=
\left({P_i\over\rho_i^2}+
      {P_j\over\rho_j^2}
\right)
\left(-\alpha\mu_{ij}+
      \beta\mu_{ij}^2\right)f_i\,,
\label{piDB}
\end{equation}
with $\alpha=\beta=1$. Here,
\begin{equation}
\mu_{ij}=
\cases{
  \medskip\displaystyle
  {({\bf v}_i-{\bf v}_j)\cdot
   ({\bf r}_i-{\bf r}_j)\over
   h_{ij}\left(|{\bf r}_i -{\bf r}_j|^2/h_{ij}^2+\eta^2\right)}
  \,,
    &if $({\bf v}_i-{\bf v}_j)\cdot({\bf r}_i-{\bf r}_j)<0$\,;\cr
  0\,,
    &if $({\bf v}_i-{\bf v}_j)\cdot({\bf r}_i-{\bf r}_j)\ge0$\,,\cr}
\label{muDB}
\end{equation}
with $\eta^2=10^{-2}$, and
the Balsara switch $f_i$ for particle $i$ is defined by
\begin{equation}
f_i={|{\bf\nabla}\cdot{\bf v}|_i\over
     |{\bf\nabla}\cdot{\bf v}|_i+
     |{\bf\nabla}\times{\bf v}|_i+
     \eta'c_i/h_i}\,,
\label{fi}
\end{equation}
with
$\eta'=10^{-5}$ preventing numerical divergences \citep{Balsara:1995JCoPh.121..357B}.  The function $f_i$
approaches unity in regions of strong compression ($|{\bf
\nabla}\cdot{\bf v}|_i>>|{\bf\nabla}\times{\bf v}|_i$) and vanishes in
regions of large vorticity ($|{\bf \nabla}\times {\bf v}|_i >>|{\bf
\nabla}\cdot {\bf v}|_i$). Consequently, our evolution equations have the advantage that the artificial
viscosity (AV) is suppressed in shear layers.  We note that the AV
term is not symmetric under interchange of the indices $i$ and $j$
(that is, $\Pi_{ij}\ne \Pi_{ji}$), because the switch $f_i$ is not
symmetrized in equation (\ref{piDB}).  Such an approach reduces
the number of
arrays shared among parallel processes.  As the term in square brackets in equation (\ref{vdottotal}) is
antisymmetric under the interchange of particles $i$ and $j$, momentum
is clearly conserved in every interaction pair.  Similarly, it is
straightforward to show total energy is conserved by our evolution
equations: $\sum_i m_i({\bf v}_i\cdot{\bf \dot v}_i
+d\Phi_i/dt+du_i/dt)=0$, where the specific internal energy
$u_i=A_i\rho_i^{\gamma-1}/(\gamma-1)$.

The evolution equations are integrated using a second-order explicit
leap-frog scheme.  For stability, the timestep must satisfy a
Courant-like condition. Specifically, we calculate the timestep as
\begin{equation}
\Delta t={\rm Min}_i
\left(\Delta t_{1,i},
            \Delta t_{2,i}
\right).
\label{good.dt}
\end{equation}
For each SPH particle $i$, we use
\begin{equation}
\Delta t_{1,i}=C_{N,1}\frac{h_i}{{\rm Max}_j\left(\Upsilon_{ij}\rho_i\right)^{1/2}}\,
\label{dt1}
\end{equation}
and
\begin{equation}
\Delta t_{2,i}=C_{N,2}
  \left({h_i\over
    \left|\dot{\bf v}_i\right|}
\right)^{1/2}\,.
\label{dt2}
\end{equation}
For the simulations presented in this paper, the Courant factors $C_{N,1}=0.4$	
and $C_{N,2}=0.05$.		
The Max$_j$ function in
equation~(\ref{dt1}) refers to the maximum of the value of its
expression for all SPH particles $j$ that are neighbors with $i$.  The
denominator of equation~(\ref{dt1}) is an approximate upper limit to
the signal propagation speed near particle $i$.

\section{SPH expression for the surface density}

We adopt a kernel function
\begin{eqnarray*}
W(r,h) = \frac{1}{\pi h^3}\left\{\begin{array}{ll}1-\frac{3}{2}\left(\frac{r}{h}\right)^2+\frac{3}{4}\left(\frac{r}{h}\right)^3, & 0\le\frac{r}{h}<1\\ \\ \frac{1}{4}\left(2-\frac{r}{h}\right)^3,& 1\le\frac{r}{h}< 2\\  \\0, & \frac{r}{h}\ge2\end{array}\right.
 = 
\left\{\begin{array}{ll}W_{out},& 0\le\frac{r}{h}< 2\\  \\0, & \frac{r}{h}\ge2\end{array}\right.
 + \left\{\begin{array}{ll}W_{in},& 0\le\frac{r}{h}< 1\\  \\0, & \frac{r}{h}\ge1\end{array}\right.,
\end{eqnarray*}
where we have defined
\begin{eqnarray*}
W_{out}&=&\frac{1}{4\pi h^3}\left(2-\frac{r}{h}\right)^3=\frac{1}{\pi h^3}\left[ 2-3\frac{r}{h}+\frac{3}{2}\left(\frac{r}{h}\right)^2-\frac{1}{4}\left(\frac{r}{h}\right)^3\right]
\\
W_{in}&=&\frac{1}{\pi h^3}\left(\frac{r}{h}-1\right)^3 = \frac{1}{\pi h^3}\left[-1+3\frac{r}{h}-3\left(\frac{r}{h}\right)^2+\left(\frac{r}{h}\right)^3\right].
\end{eqnarray*}
The function $W_{out}$ is just the kernel function in the outer regions of the compact support ($1\le r/h<2$), while $W_{in}$ is the difference between the kernel function in the inner and outer domains.
The distance $r$ from the center of an SPH particle to the points on a line passing vertically with horizontal offset $\rho$ from that center is given by $r=\sqrt{Z^2+\rho^2}$, where $Z$ is the vertical offset.
Thus, we may define $z_{\rm out} = \sqrt{4h^2-\rho^2}$, and if $\rho<h$ the quantity $z_{\rm in} = \sqrt{h^2-\rho^2}$, to define the integration bounds for the kernel-based  surface density
\begin{eqnarray}
\Sigma &=& \sum_i m_i \int W(r,h_i) dZ.
\end{eqnarray}
In particular, a particle $i$ that is passed through in only the outer part of its kernel by the line being integrated along 
(so $h_i<\rho<2h_i$) will contribute 
\begin{eqnarray}
\Sigma_{\rm out, i} &=& \frac{m_i}{\pi h_i^2}\left[2i_0\left(\frac{z_{out}}{h}\right) -3i_1\left(\frac{z_{out}}{h}\right) +\frac{3}{2}i_2\left(\frac{z_{out}}{h}\right) -\frac{1}{4}i_3\left(\frac{z_{out}}{h}\right) \right]
\end{eqnarray}
to the surface density.  While if $\rho<h_i$ then particle $i$ contributes
\begin{eqnarray}
\Sigma_{\rm in,i} &=& \Sigma_{\rm out, i}  +\frac{m_i}{\pi h_i^2}\left[-i_0\left(\frac{z_{in}}{h}\right) +3i_1\left(\frac{z_{in}}{h}\right) -3i_2\left(\frac{z_{in}}{h}\right) +i_3\left(\frac{z_{in}}{h}\right) \right].
\end{eqnarray}
Here we make use of the vertical symmetry of the kernel function to define the following integrals:
\begin{eqnarray*}
I_0(z)=hi_0\left(\frac{z}{h}\right) &=& \int_{-z}^z ~dZ= 2z\\
I_1(z) =h^2i_1\left(\frac{z}{h}\right)&=&\int_{-z}^z r(Z)~dZ = \int_{-z}^z\sqrt{Z^2+\rho^2}~dZ =  \left.\frac{1}{2}\left(Z\sqrt{Z^2+\rho^2}+\rho^2\ln\left[Z+\sqrt{Z^2+\rho^2}\right]\right)\right|_{-z}^z\\&=&z\sqrt{z^2+\rho^2}+\rho^2\ln\left(\frac{z+\sqrt{z^2+\rho^2}}{\rho}\right)\\
I_2(z)=h^3i_2\left(\frac{z}{h}\right)&=&\int_{-z}^z r^2dZ = \int_{-z}^z(Z^2+\rho^2)dZ = \frac{2z}{3}(z^2+3\rho^2)\\
I_3(z)=h^4i_3\left(\frac{z}{h}\right)&=&\int_{-z}^z r^3dZ= \int_{-z}^z(Z^2+\rho^2)^{3/2}dZ =  \frac{1}{4}\left(z\sqrt{z^2+\rho^2}\left[2z^2+5\rho^2\right]+3\rho^4\ln\left[\frac{z+\sqrt{z^2+\rho^2}}{\rho}\right]\right).\end{eqnarray*}

\end{appendix}
\bibliographystyle{apj}

\end{document}